\documentclass[preprint,review,12pt,a4paper,3p]{elsarticle}
\usepackage[version=3]{mhchem} 
\usepackage[]{geometry}
\usepackage[scriptsize,tight]{subfigure}
\usepackage{graphicx}
\usepackage{multirow}
\usepackage{dcolumn}
\usepackage{bm}
\usepackage{color,soul}
\usepackage[mathlines]{lineno}
\usepackage{hyperref}

\biboptions{longnamesfirst,sort&compress,semicolon}

\begin{document}

\begin{frontmatter}

\title{Electronic and structural properties of M\"obius boron--nitride and
carbon nanobelts}

\author{C. Aguiar$^{1}$}
\author{N. Dattani$^{2,3}$}
\ead{nike@hpqc.org}
\author{I.~Camps\corref{coric}$^{1,3}$}
\ead{icamps@unifal-mg.edu.br}
\cortext[coric]{Corresponding authors}

\address{$^{1}$Laborat\'orio de Modelagem Computacional - \emph{La}Model,
Instituto de Ci\^{e}ncias Exatas - ICEx. Universidade Federal de Alfenas -
UNIFAL-MG, Alfenas, Minas Gerais, Brasil}
\address{$^{2}$HPQC College, Waterloo, Canada}
\address{$^{3}$HPQC Labs, Waterloo, Canada}

\begin{abstract}
Using the semiempirical tight binding method as implemented in the xTB program,
we characterized M\"obius boron–nitride and carbon-based nanobelts with
different
sizes and compared them with normal nanobelts. The calculated properties
include the infrared spectra, the highest occupied molecular orbital (HOMO),
the lowest unoccupied molecular orbital (LUMO), the energy gap, the chemical
potential, and the molecular hardness. The agreement between the peaks
positions from
theoretical infrared spectra compared with experimental ones for all systems,
validate the used methodology. Our findings show that for the boron–nitride
based nanobelts, the calculated properties have opposite monotonic relationship
with the size of the systems whereas, for the carbon-based, the properties show
the same monotonic relationship for both types of nanobelts. Also, the torsion
presented on the M\"obius nanobelts, in the case of boron–nitride, induced an
inhomogeneous surface distribution for the HOMO orbitals. In all cases, the
properties vary with the increase in the size of the nanobelts indicating that
it is possible to choose the desired values by changing the size and type of
the systems.
\end{abstract}

\begin{keyword}
nanotechnology \sep nanobelts \sep boron nitride \sep carbon
\end{keyword}

\end{frontmatter}

\section{Introduction}
\label{Sec:Intro}
Carbon (C) can present three different forms of hybridization ($sp$, $sp^2$ and
$sp^3$) and the way in which they can combine with other elements is the basis
of countless
researches~\cite{Kasalkova-Nanomaterials-11-2368-2021,Pedrosa-MaterialsResearch-23-20190493-2020}.

The combination of this element on a nanometric
scale gave rise to structures called
nanocarbons~\cite{Jirimali-Materials-15-3969-2022}. They are characterized by
showing different geometric and dimensional configurations, such as fullerenes,
carbon nanotubes, graphene nanoribbons, graphene oxides and
nanodiamonds~\cite{Kasalkova-Nanomaterials-11-2368-2021,Segawa-Science-365-272-2019}.

Such structures have attracted interest for promising applications in
nanobiomedicine~\cite{Cheung-Chem.Eur.J.-26-14791-2020,Panwar-Chem.Rev.-119-9559-2019}
 and optoelectronics~\cite{Saba_2019,Shearer-275-2020}. According to the size
and/or
different topologies these nanomaterials cause different reactivity in contact
with other
materials~\cite{Pedrosa-MaterialsResearch-23-20190493-2020,Itami-NanoLett.-20-4718-2020},
favored or not, according to the geometry of the
nanomaterial, which varies according to the increase in the pyramidalization
angle and misalignment of the $\pi$ orbitals between the C
atoms~\cite{Pedrosa-MaterialsResearch-23-20190493-2020}. Thus,
different syntheses and functionalization have been carried out to obtain
different nanostructures to achieve greater compatibility with other materials,
seeking to improve and expand nanotechnological
applications~\cite{Cheung-Chem.Eur.J.-26-14791-2020,Itami-NanoLett.-20-4718-2020,Guo-Nat.Chem.-13-402-2021,Yang-Chem.Rev.-120-2693-2020}.

Povie et al., made a breakthrough with the
synthesis of carbon nanobelts
(CNBs), whose simple structure in the form of a ring or belt generates two
faces, one internal and one external, not convertible to each
other~\cite{Povie-Science-356-172-2017}.
Carbon nanobelts represent segments of single walled carbon nanotubes
containing a benzene ring cycle with $p$ orbitals aligned in a
plane~\cite{Xia-Angew.Chem.-133-10399-2021}. Such
behavior allows them to be classified as armchair, zigzag or chiral nanoribbons
according to the chirality
index~\cite{Xia-Angew.Chem.-133-10399-2021,Price-NatureSynthesis-1-502-2022}.
In addition to the ring shape, carbon
nanobelts are attractive molecules due to their synthetic challenges and
differentiated
properties~\cite{Lu-Chem-2-619-2017,Chen-J.Phys.Org.Chem.-33--2020}. Its
formulation covers concepts of
conjugation, aromaticity and strain, also providing important information on
chirality and bottom-up synthesis of C nanotubes, which continues to be a
challenge and has caused limitations in its
applications~\cite{Segawa-NatureSynthesis-1-535-2022}. For the
synthesis of nanobelts, three steps must be considered, the first consists of
macrocyclization, from carbon sources, the second must occur with the formation
of the belt in order to create the double-stranded structure, and finally the
induction step stress required to bend the $sp^^2$ hybridized carbon skeleton
into
a cylindrical
topology~\cite{Povie-Science-356-172-2017,Li-Acc.Mater.Res.-2-681-2021}.

But recently, Segawa et al., obtained a new structure, which are
structurally uniform, and can be obtained by the bottom-up method in 14 steps
of synthesis~\cite{Segawa-NatureSynthesis-1-535-2022}. These are M\"obius
carbon
nanobelts (MCNBs), whose CNBs
structure, when subjected to torsion, should manifest different properties and
molecular movements when compared to nanocarbons with a common belt
topology~\cite{Povie-Science-356-172-2017}. Density functional theory (DFT)
calculations show that MCNBs have a
higher strain energy than CNBs of the same
size~\cite{Segawa-NatureSynthesis-1-535-2022}. However, producing
torsion in CNBs can be difficult to control as strain energy is the major
obstacle in the synthesis of MCNBs~\cite{Povie-Science-356-172-2017}. For this,
saturated ligands (–CH\textsubscript{2}O–)
or chalcogen atom ligands (–S–) are used to reduce and control the stress
caused by the M\"obius
shape~\cite{Nishigaki-J.Am.Chem.Soc.-141-14955-2019,Wang-Angew.Chem.Int.Ed.-60-18443-2021}.
 Nonetheless, calculations of strain
energies
showed that MCNBs are synthetically accessible and that strain decreases with
increasing MCNBs size~\cite{Ajami-Nature-426-819-2003}. Data from nuclear
magnetic resonance spectroscopy
and theoretical calculations show that the torsion structure of the M\"obius
band
moves rapidly in solution~\cite{Segawa-NatureSynthesis-1-535-2022}.
Furthermore, chirality arising from the M\"obius
structure has been demonstrated experimentally using chiral separation by high
performance liquid chromatography (HPLC) and circular dichroism (DC)
spectroscopy~\cite{Segawa-NatureSynthesis-1-535-2022}. Besides, spectroscopy
data on excitation at 380 nm showed
blue-green fluorescence beyond 10\% quantum
yield~\cite{Segawa-NatureSynthesis-1-535-2022}.

Given the above, new synthetic route strategies, different topologies and
varied functionalization, combined with deformation calculations may
contribute to the improvement of new nanocarbon materials from CNBs, including
M\"obius. This would make it possible to properly relate structure and function
for applications in several areas. Thus, the objective of this work is to
determine the electronic and structural properties of carbon and boron--nitride
nanobelts and M\"obius nanobelts with different sizes. The knowledge of such
properties can help to develop sensors and filters for heavy metal,
green house and hazardous gases.

\section{Materials and Methods}
\label{Sec:Method}

In this work two type of nanobelts were used for C and B--N systems. The
first one consist in nanobelts whereas the second one are M\"obius nanobelts
(twisted
nanobelts). The structures were generated starting with a cell with 2 units of
(10,0)
nanosheet repeated N times (10, 15, 20, 25 and 30) in the z direction and then
wrapped $360^o$. After that, the periodicity was removed and the atoms of the
borders
were
passivated with Hydrogen.
In case of M\"obius nanobelts, after repetition, the nanobelts were twisted
$180^o$ and then wrapped. All the structures were generated using the
Virtual NanoLab Atomistix
ToolKit software~\cite{VNL}.

The nomenclature to identify the systems is described as follows.
BNNB\textsubscript{N} (CNB\textsubscript{N}) for boron--nitride (carbon)
nanobelts and MBNNB\textsubscript{N}
(MCNB\textsubscript{N}) for M\"obius boron--nitride (carbon) nanobelt. In all
cases, N indicates the number of repetitions.

Using the semiempirical tight binding method as implemented in the xTB
program~\cite{xTB_1}, the electronic and structural parameters were
calculated. The calculations were done using the GFN2--xTB method, an accurate
self--consistent
method that includes multipole electrostatics and density--dependent dispersion
contributions~\cite{xTB_3} with the extreme optimization level that ensures
convergence energy of $5\times10^{-8}$~E\textsubscript{h} and gradient norm
convergence of
$5\times10^{-5}$~E\textsubscript{h}/a\textsubscript{0} (a\textsubscript{0} is
the Bohr
radii).

For each optimized structure, the highest occupied
molecular orbital, HOMO ($\varepsilon_H$); the lowest unoccupied
molecular orbital, LUMO ($\varepsilon_L$); the energy gap ($\Delta
\varepsilon$) between HOMO and LUMO orbitals ($\Delta \varepsilon =
\varepsilon_H -
\varepsilon_L$); the chemical potential ($\mu$); the molecular hardness
($\eta$); and the infrared spectra were determined.

Considering the approximation that ignores orbital relaxation after an
electron is removed from the system (Koopmans\textquoteright~s
theorem~\cite{koopmans-Physica-1-104-1933,luo-J.Phys.Chem.A-110-12005-2006,salzner-J.Chem.Phys.-131-231101-2009,tsuneda-J.Chem.Phys.-133-174101-2010}
together with Janak\textquoteright~s
theorem~\cite{janak-Phys.Rev.-1978-7165-18}) it is possible to estimate the
chemical potential ($\mu$), the molecular hardness
($\eta$)~\cite{zhan-J.Phys.Chem.A-107-4184-2003}, and the electrophilicity
index ($\omega$)~\cite{parr-J.Am.Chem.Soc.-121-1922-1999} from the HOMO and
LUMO energies $\epsilon_{H}$ and $\epsilon_{L}$ as follows:

\begin{eqnarray}
\label{eq:param}
\mu  \cong \frac{{{\varepsilon _L} + {\varepsilon _H}}}{2}\\
\eta  \cong \frac{{{\varepsilon _L} - {\varepsilon _H}}}{2}\\
\omega  = \frac{{{\mu ^2}}}{{2\eta }}
\end{eqnarray}

\section{Results and discussion}
\label{Sec:Results}

Figures~\ref{Fig:StructALL_BN} and~\ref{Fig:StructALL_CC} shown the top view
of the optimized structures. Panel A shows the nanobelts and panel B shows
the M\"obius nanobelts. With the increase of repetitions (n), the diameter
increase too, starting from $13.81$~\AA~($13.77$~\AA) to
$40.79$~\AA~($40.77$~\AA) for boron--nitride (carbon) nanobelts. The minimum
number of repetition used was 10 to avoid stressed structures.

\newcommand{\sizeA}{7.5cm}
\begin{figure}[tbph]
\centering
\includegraphics[width=\sizeA,angle=-90]{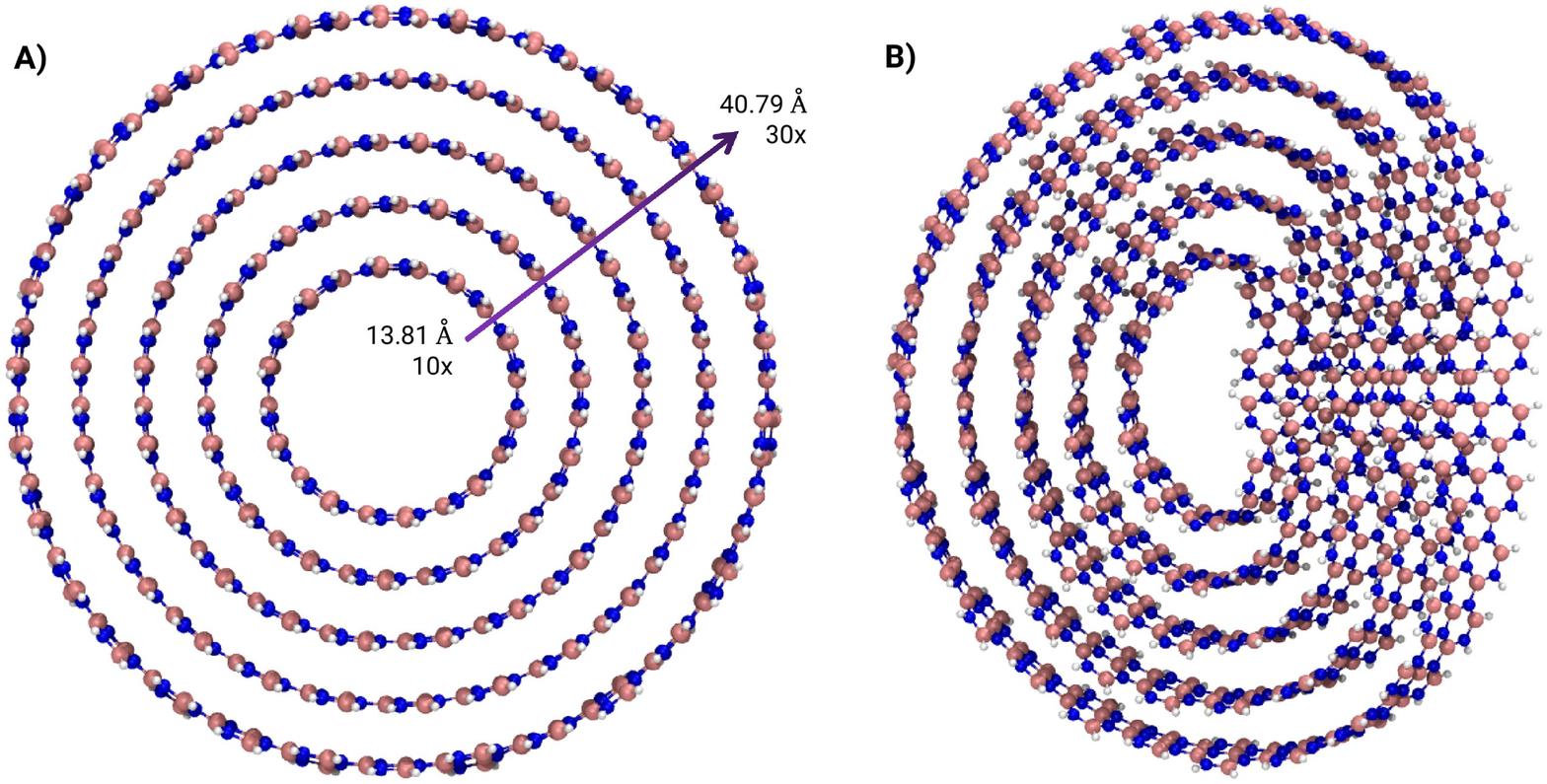}
\caption{\label{Fig:StructALL_BN} Top view of: A) Boron--nitride nanobelts
(BNNB) with
minimum/maximum diameter and repetition. B)
M\"obius boron--nitride nanobelts (MBNNB). Image rendered with VMD~\cite{vmd}
software.}
\end{figure}

\begin{figure}[tbph]
\centering
\includegraphics[width=\sizeA,angle=-90]{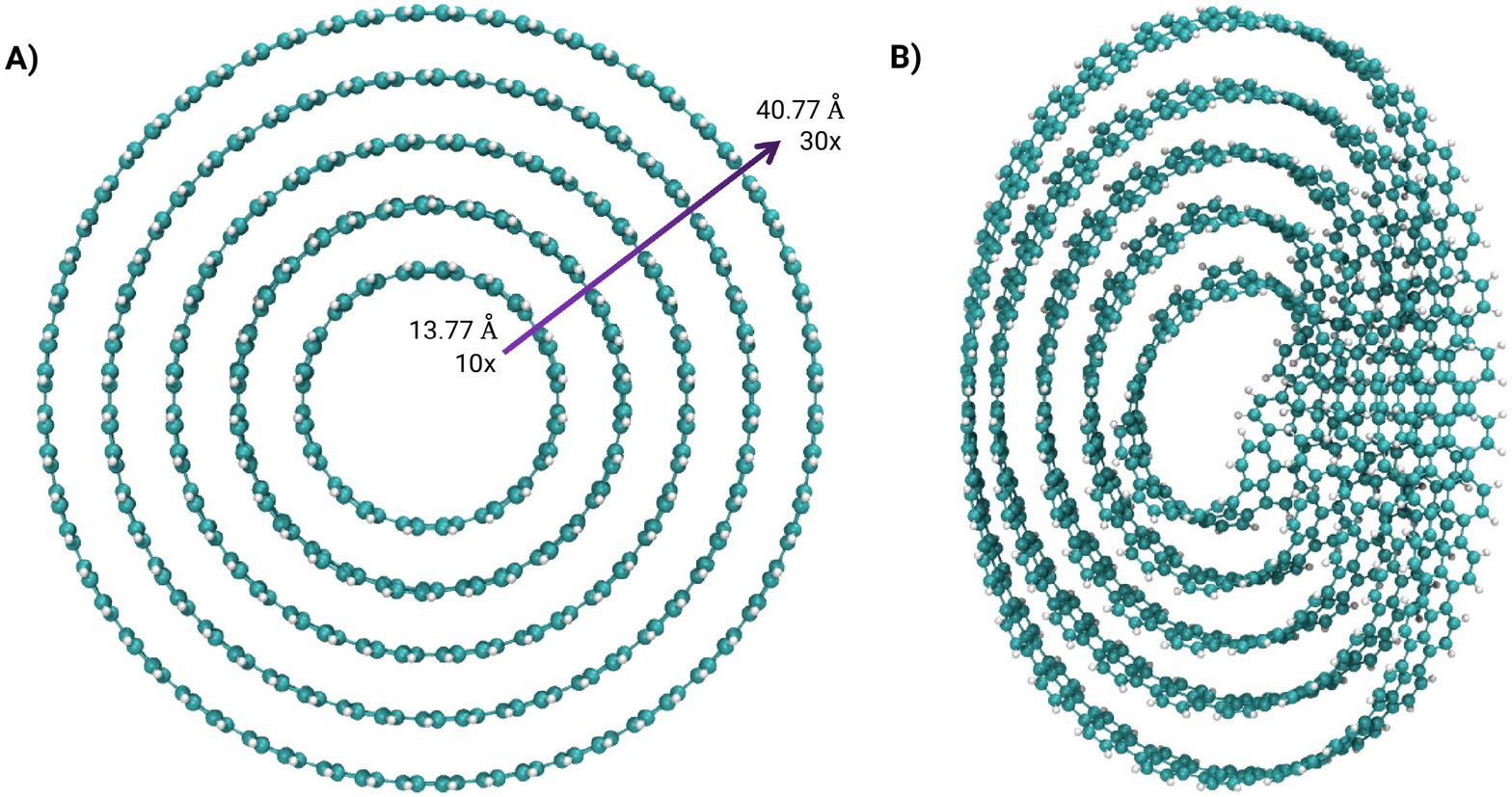}
\caption{\label{Fig:StructALL_CC} Top view of: A) Carbon nanobelts (CNB) with
minimum/maximum diameter and repetition. B)
M\"obius carbon nanobelts (MCNB). Image rendered with VMD~\cite{vmd} software.}
\end{figure}

The calculated infrared spectra for each system are shown in
figures~\ref{Fig:IR-BN} and~\ref{Fig:IR-C}. At first sight
(figure~\ref{Fig:IR-BN}), the spectra for both
systems, BNNB and MBNNB are very similar, indicating that the torsion on the
M\"obius nanobelts did not appreciable change the principal oscillation modes.
In both cases, the B--N stretching
(in--plane and out--of--plane) and radial R mode (out--of--plane buckling) are
observed. The oscillations around $800$~cm\textsuperscript{-1} correspond to
the out--of--plane buckling mode. All the resonances in the high--frequency
regime above $1200$~cm\textsuperscript{-1} consists of transverse optical (T)
and
longitudinal optical (L) phonon modes. Oscillations around
$1200$~cm\textsuperscript{-1} and $1380$~cm\textsuperscript{-1}
correspond to bond--bending or T modes and around $1340$~cm\textsuperscript{-1}
and $1420$~cm\textsuperscript{-1} correspond to bond--stretching or L
modes~\cite{Wirtz-Phys.Rev.B-68-45425-2003,Singh-Sci.Rep.-6-35535-2016,
Harrison-NanoscaleAdv.-1-1693-2019,Zhang-Catalysts-10-596-2020}.

The case for the carbon nanobelts systems, CNB and MCNB, is different. For both
systems, the fingerprint region ($600-1500$~cm\textsuperscript{-1}) is visible
but the peaks are with very dissimilar intensities, being greater for the MCNB
structures. For carbon based organic systems, the resonances around
$900-675$~cm\textsuperscript{-1} consist on out--of--plane C--H oscillation,
around $1500-1400$~cm\textsuperscript{-1} were identified as C--C stretch
(in--ring modes), and around $3100-3000$~cm\textsuperscript{-1}, as C--H bond
stretch~\cite{Nandiyanto-Indones.J.Sci.Technol.-4-97-2019}. These results
indicate that the MCNB are more freely to oscillate than the CNB.

\renewcommand{\sizeA}{8.0cm}
\begin{figure}[tbph]
\centering
\subfigure[][BNNB
system]{\includegraphics[width=\sizeA]{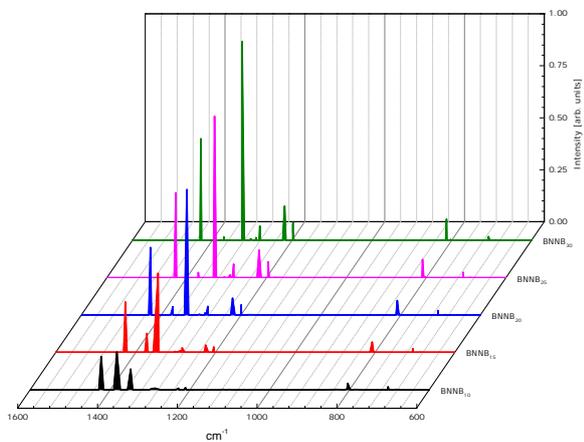}
\label{Fig:IR-BNNB}}
\subfigure[][MBNNB
system]{\includegraphics[width=\sizeA]{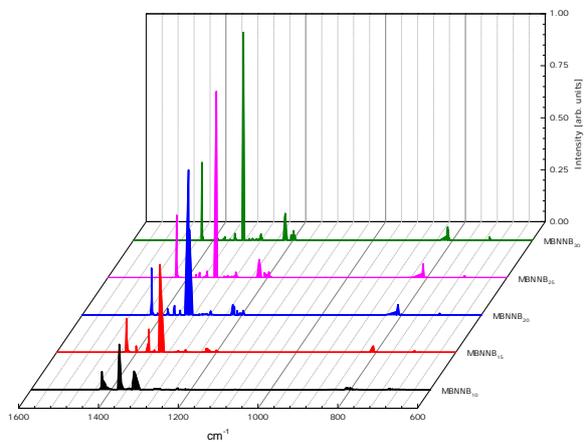}
\label{Fig:IR-MBNNB}}
\caption{\label{Fig:IR-BN} Calculated infrared spectra for boron--nitride
nanobelts.}
\end{figure}

\renewcommand{\sizeA}{8.0cm}
\begin{figure}[tbph]
\centering
\subfigure[][CNB system]{\includegraphics[width=\sizeA]{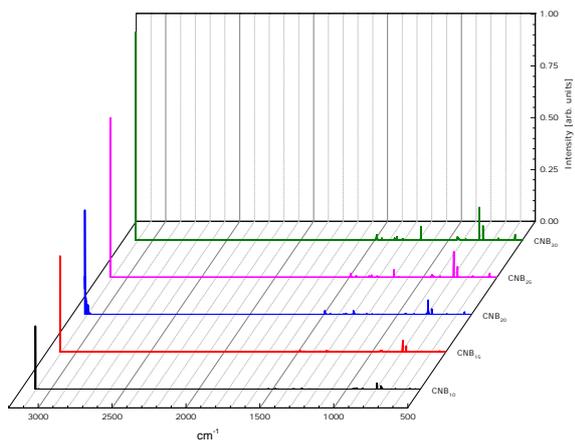}
\label{Fig:IR-CNB}}
\subfigure[][MCNB system]{\includegraphics[width=\sizeA]{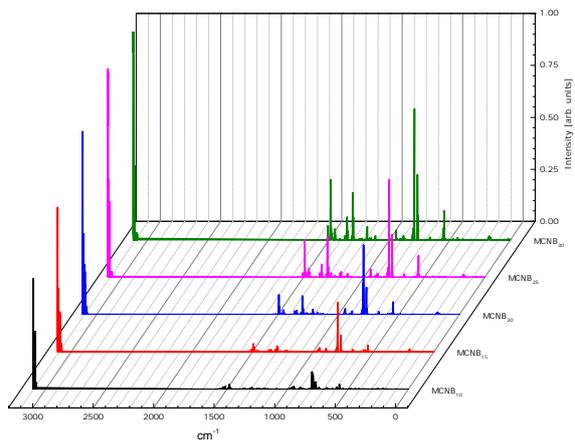}
\label{Fig:IR-MCNB}}
\caption{\label{Fig:IR-C} Calculated infrared spectra for carbon nanobelts.}
\end{figure}

The electronic properties for boron--nitride and carbon based systems are shown
in figures~\ref{Fig:ELECT_PROP-BN}.
Figures~\ref{Fig:HOMO-BN} and~\ref{Fig:LUMO-BN} show the energies of HOMO
and LUMO boundary orbitals. To some extent, from the energies of these
orbitals, it is possible to know how reactive the system is. The
electron--donor character (electron--donor capacity) is measured by the HOMO
energy whereas the electron--acceptor character (resistance to accepting
electrons) is measured by the LUMO energy. From these figures, we can see that,
in case of HOMO energy, the BNNB and MBNNB have opposite behavior. With the
increase of the number of repeat units, the capacity to donate electrons is
decreased for the MBNNB system and increased for the BNNB. On the other hand,
the behavior of LUMO energy with the increase of systems size is similar,
increasing the resistance to accept electrons.

\renewcommand{\sizeA}{5.0cm}
\begin{figure}[tbph]
\centering
\subfigure[][HOMO energy]{\includegraphics[width=\sizeA]{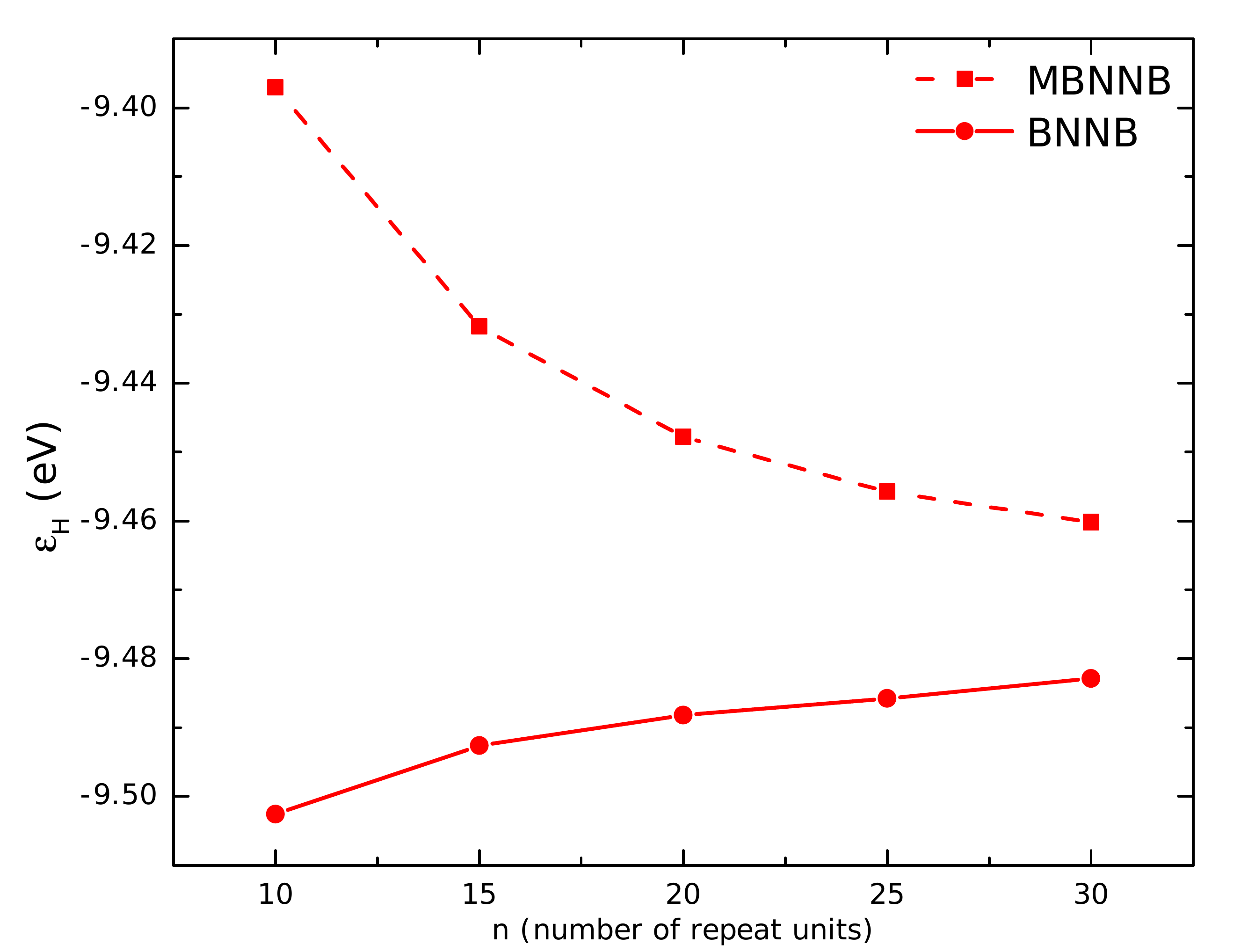}
\label{Fig:HOMO-BN}}
\subfigure[][LUMO energy]{\includegraphics[width=\sizeA]{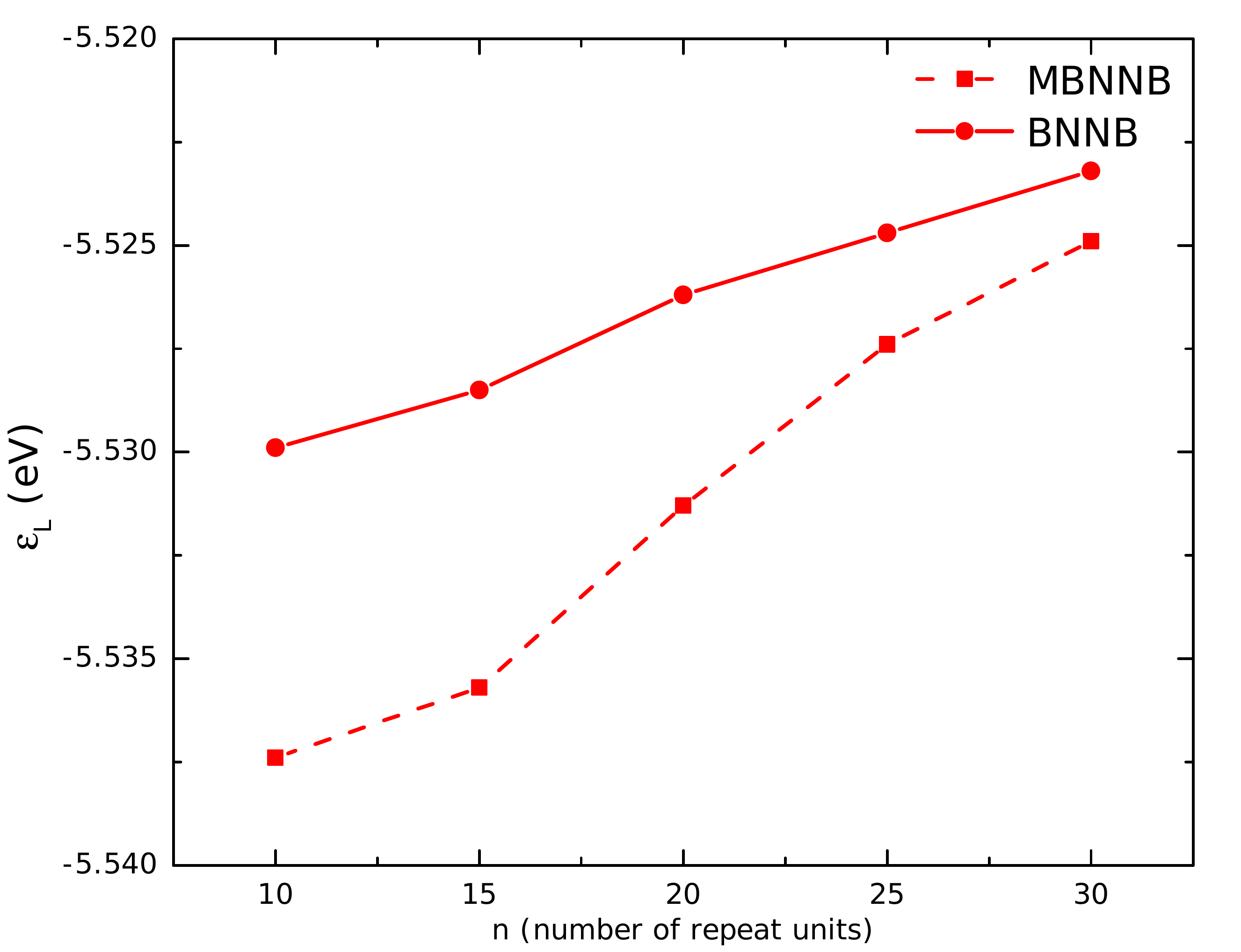}
\label{Fig:LUMO-BN}}
\subfigure[][Energy gap]{\includegraphics[width=\sizeA]{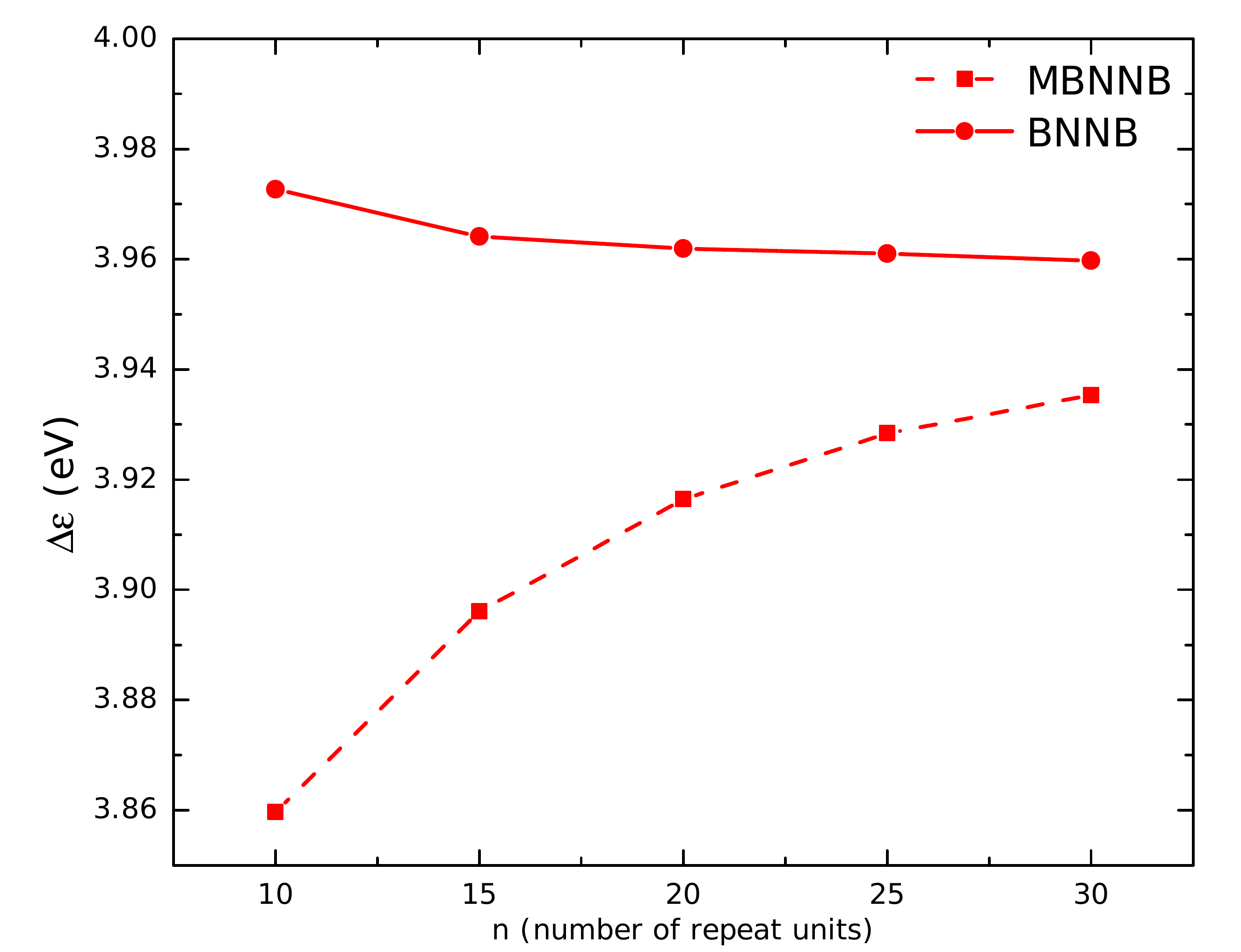}
\label{Fig:Gap-BN}}\\
\subfigure[][Chemical potential]{\includegraphics[width=\sizeA]{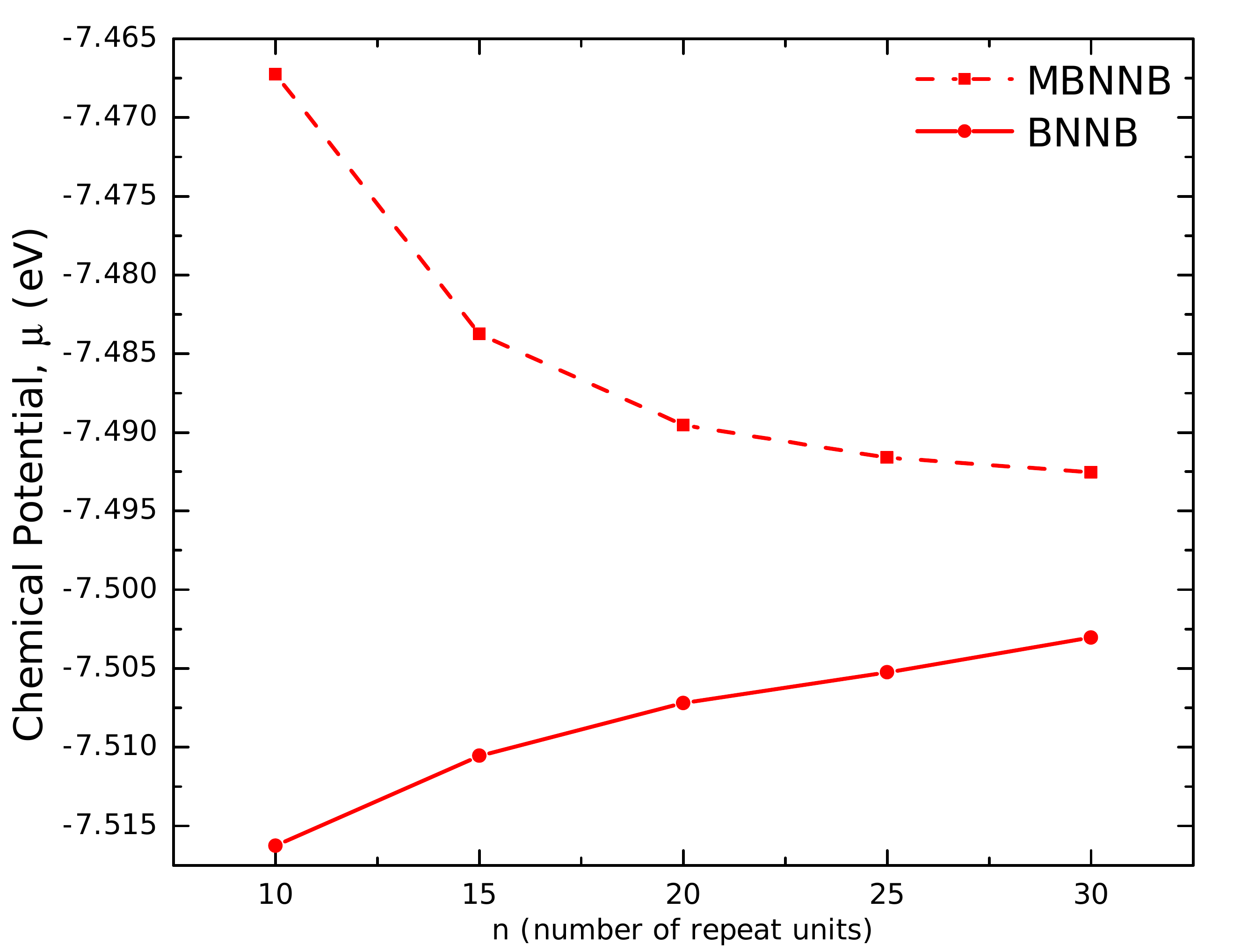}
\label{Fig:ChemicalPotential-BN}}
\subfigure[][Molecular Hardness]{\includegraphics[width=\sizeA]{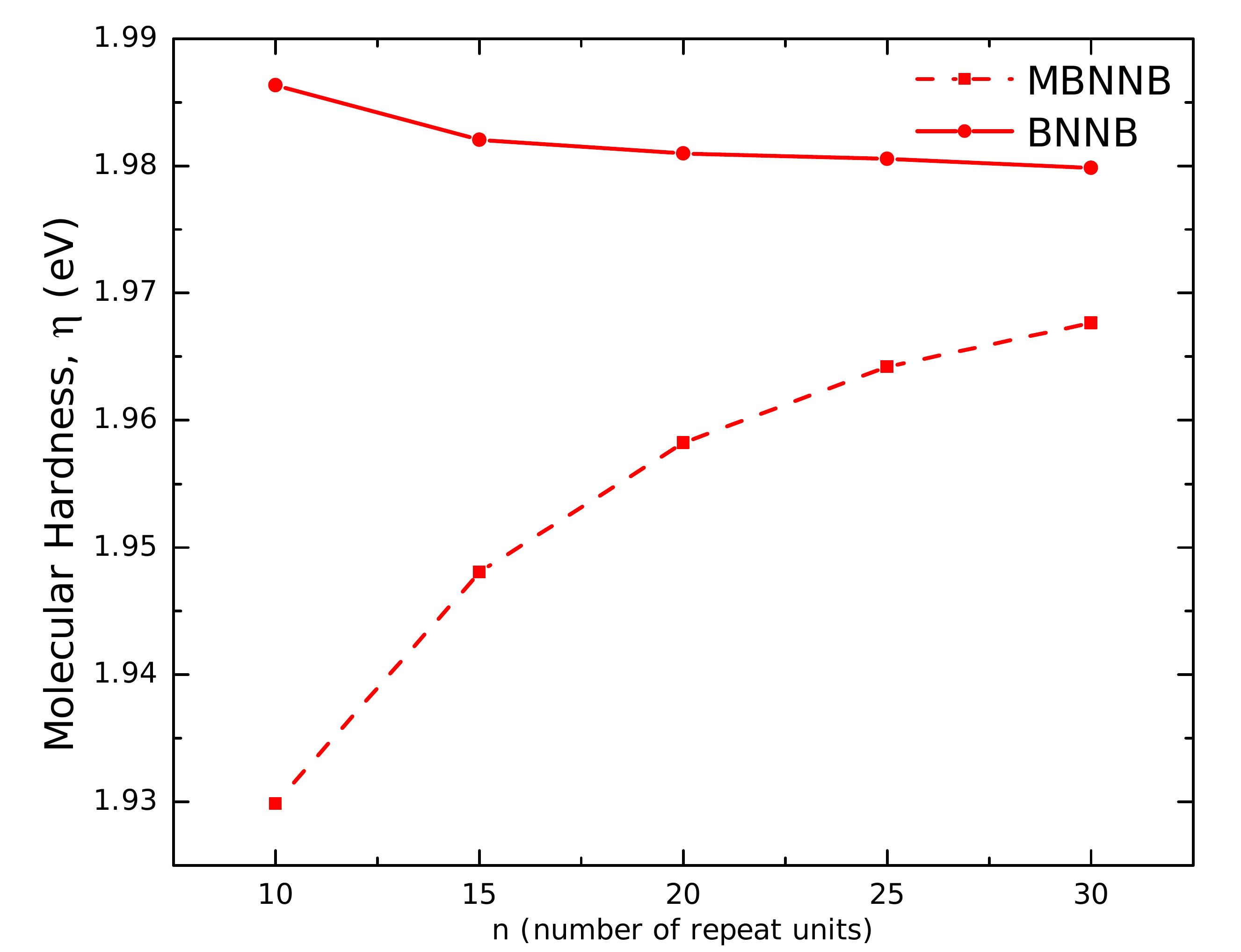}
\label{Fig:MolecularHardness-BN}}
\subfigure[][Electrophilicity
Index]{\includegraphics[width=\sizeA]{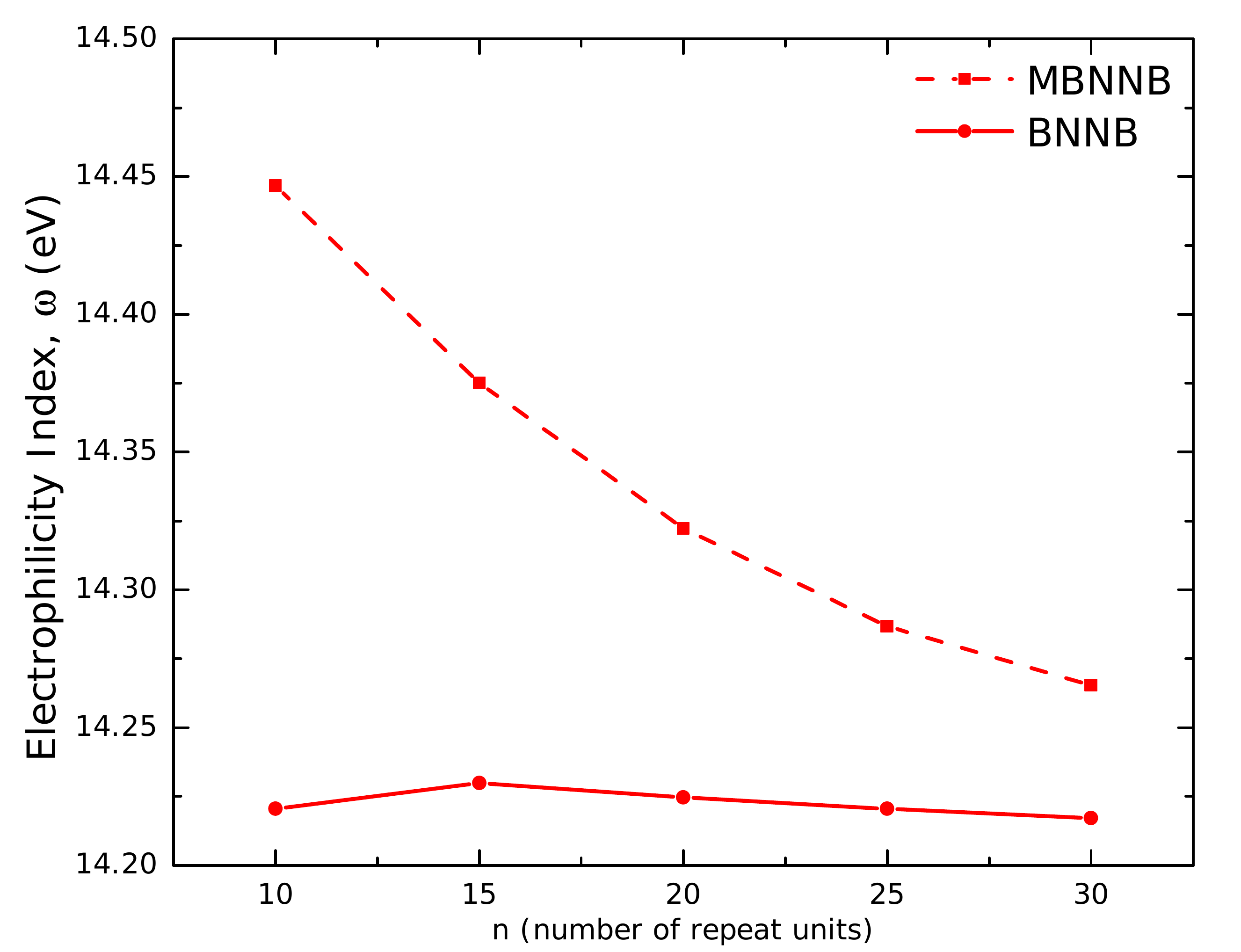}
\label{Fig:Electrophilicity-BN}}
\caption{\label{Fig:ELECT_PROP-BN} Calculated electronic properties for
boron--nitride nanobelts.}
\end{figure}

Top view of the 3--D diagrams of the HOMO and LUMO surfaces for the
BNNB\textsubscript{20} and MBNNB\textsubscript{20} are shown in
figures~\ref{Fig:OrbitalsBN20} (the 3--D
diagrams for all the
structures are shown in figures~\ref{FigS:OrbitalsBNNB} and
\ref{FigS:OrbitalsMBNNB} of Supplementary material). The distribution
of HOMO and LUMO surfaces for the BNNB are, as expected, in accordance with the
symmetry of the system being distributed homogeneously over all the structure.
In case of the MBNNB, the torsion on the nanobelt induced an
strain on the structure that in turn, modify how the orbitals are distributed.
This modification is stronger for the HOMO where the orbital volume
is redistributed having regions with smaller/greater volumes. As the volume of
the orbitals are proportional to the
probability to find the electrons, the electron--donor regions changed too,
with low/high localized HOMO zones. On the other hand, the LUMO surface shows
very little inhomogeneities. This behavior is the same for all the MBNNB as
shown in figure~\ref{FigS:OrbitalsMBNNB}.

\renewcommand{\sizeA}{3.5cm}
\begin{figure}[tbph]
\centering
\begin{tabular}{ccc}
\subfigure[BNNB\textsubscript{20}
structure]{\includegraphics[width=\sizeA]{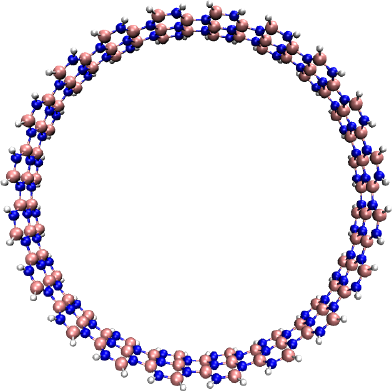}} &
\subfigure[BNNB\textsubscript{20}
HOMO]{\includegraphics[width=\sizeA]{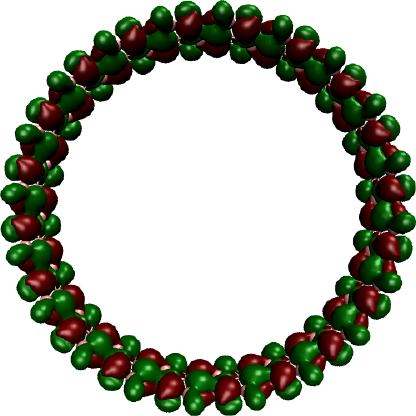}} &
\subfigure[BNNB\textsubscript{20}
LUMO]{\includegraphics[width=\sizeA]{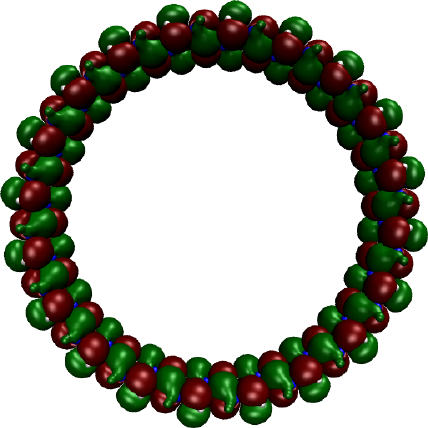}} \\

\subfigure[MBNNB\textsubscript{20}
structure]{\includegraphics[width=\sizeA]{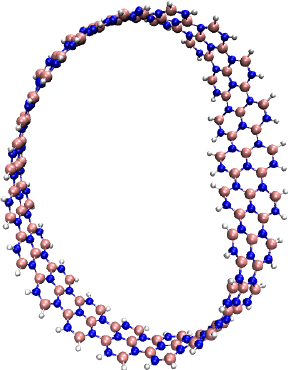}} &
\subfigure[MBNNB\textsubscript{20}
HOMO]{\includegraphics[width=\sizeA]{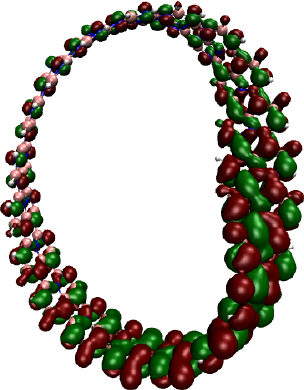}} &
\subfigure[MBNNB\textsubscript{20}
LUMO]{\includegraphics[width=\sizeA]{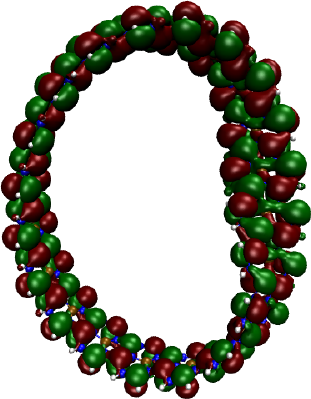}} \\

\end{tabular}
\caption{\label{Fig:OrbitalsBN20} Structures, HOMO and LUMO surfaces for
BNNB\textsubscript{20} and MBNNB\textsubscript{20} systems. Red (green) color
represents negative (positive) values. Orbital surfaces rendered with with
isovalue equal to $0.001$ and with VMD~\cite{vmd} software.}
\end{figure}

The gap ($\Delta \varepsilon$), chemical potential ($\mu$), molecular
hardness ($\eta$), and  electrophilicity index ($\omega$) can be used to
estimate the chemical reactivity and hardness of the system together with its
molecular stability. Molecules with
high (low) gap are those with high (low) molecular
stability~\cite{zhang-J.Phys.Chem.A-111-1554-2007}. Positive values for $\eta$
indicates that the redistribution of electrons in the molecule is energetically
unfavorable. Also, the higher $\eta$ value is, the more chemically stable the
molecule is and, therefore, harder the rearrangement of its
electrons~\cite{parr-J.Am.Chem.Soc.-105-7512-1983}. The electrophilicity
index ($\omega$), can be used as a measure of the molecule energy lowering due
to the maximal electron flow between the environment and the
molecule~\cite{parr-J.Am.Chem.Soc.-121-1922-1999}.

Comparing the values of
$\Delta \varepsilon$, $\mu$, $\eta$ and $\omega$ for BNNB and MBNNB systems, we
can see that the values are on the same order of magnitude. The main difference
is in how these properties change with the increase of the number of repeated
units used to build the nanobelts. These can be used to design structures
with a fine control of these properties.

\renewcommand{\sizeA}{5.0cm}
\begin{figure}[tbph]
\centering
\subfigure[][HOMO energy]{\includegraphics[width=\sizeA]{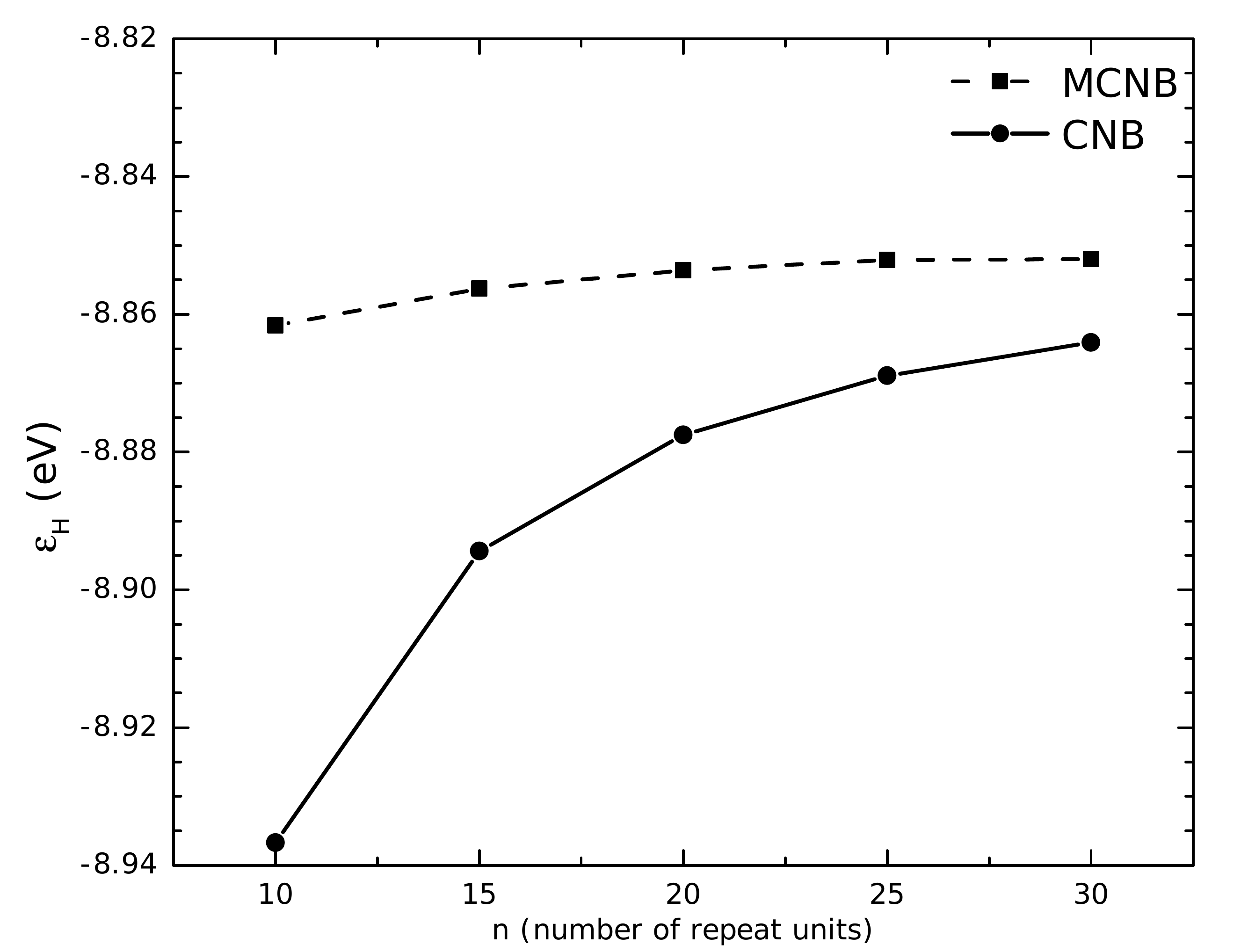}
\label{Fig:HOMO-C}}
\subfigure[][LUMO energy]{\includegraphics[width=\sizeA]{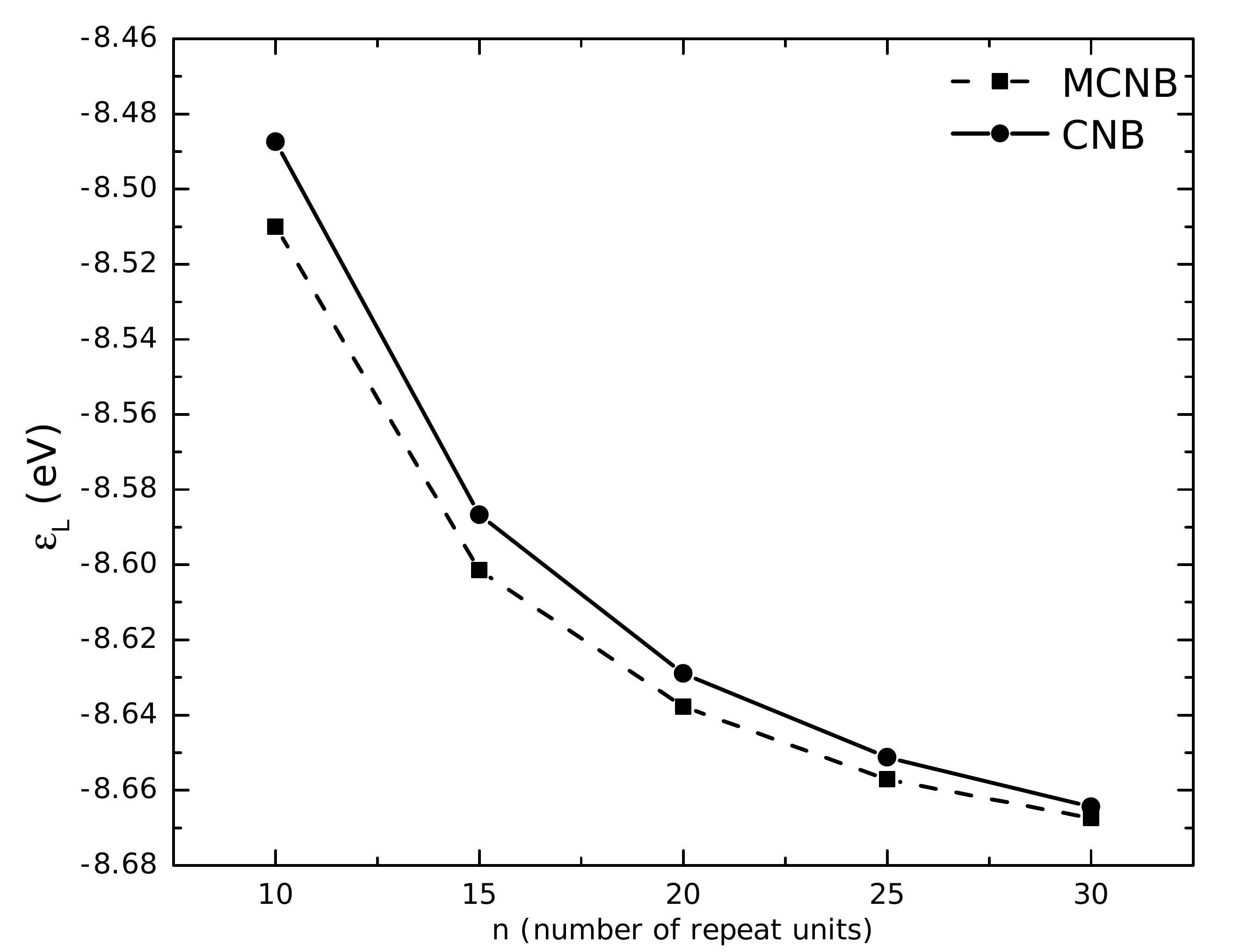}
\label{Fig:LUMO-C}}
\subfigure[][Energy gap]{\includegraphics[width=\sizeA]{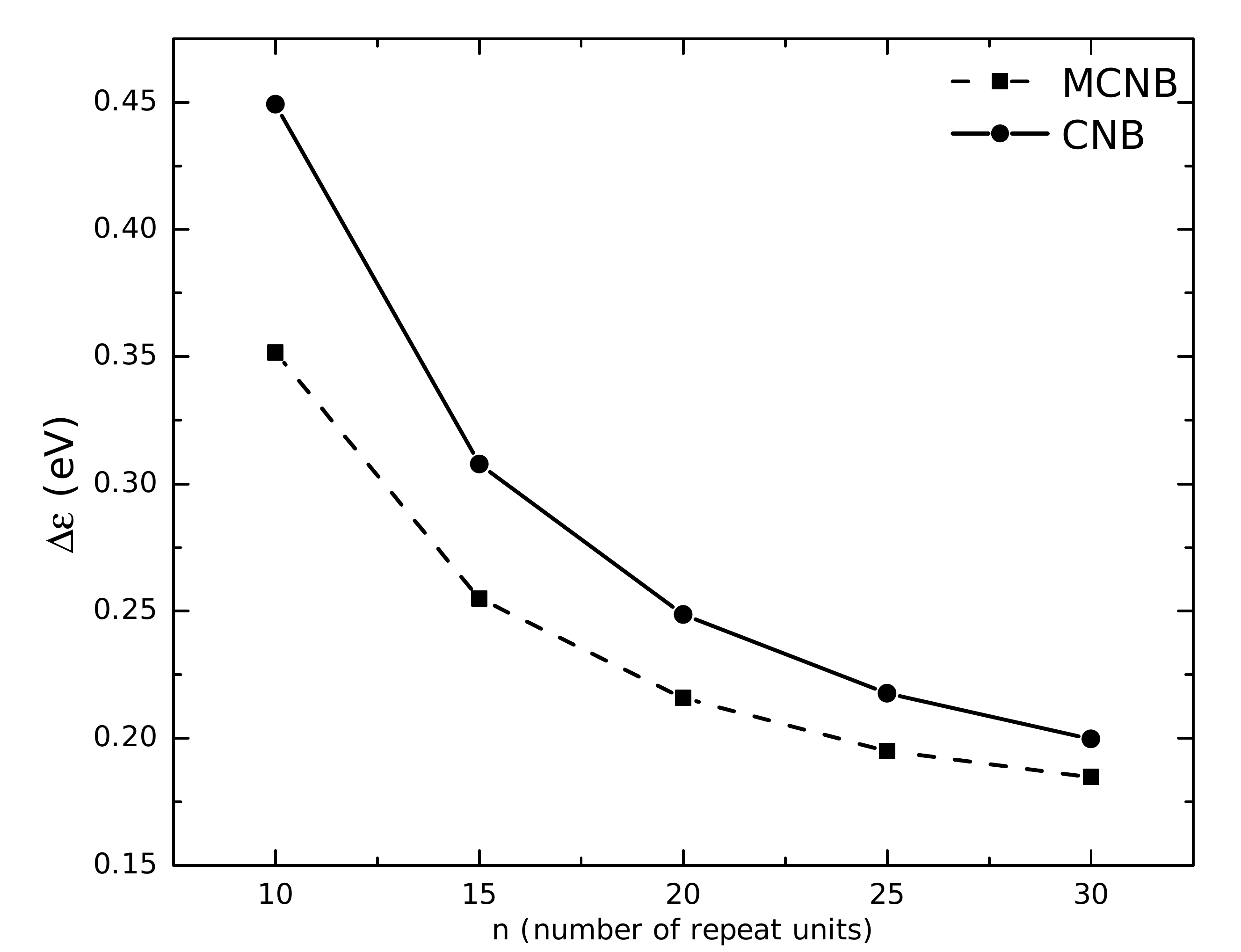}
\label{Fig:Gap-C}}\\
\subfigure[][Chemical potential]{\includegraphics[width=\sizeA]{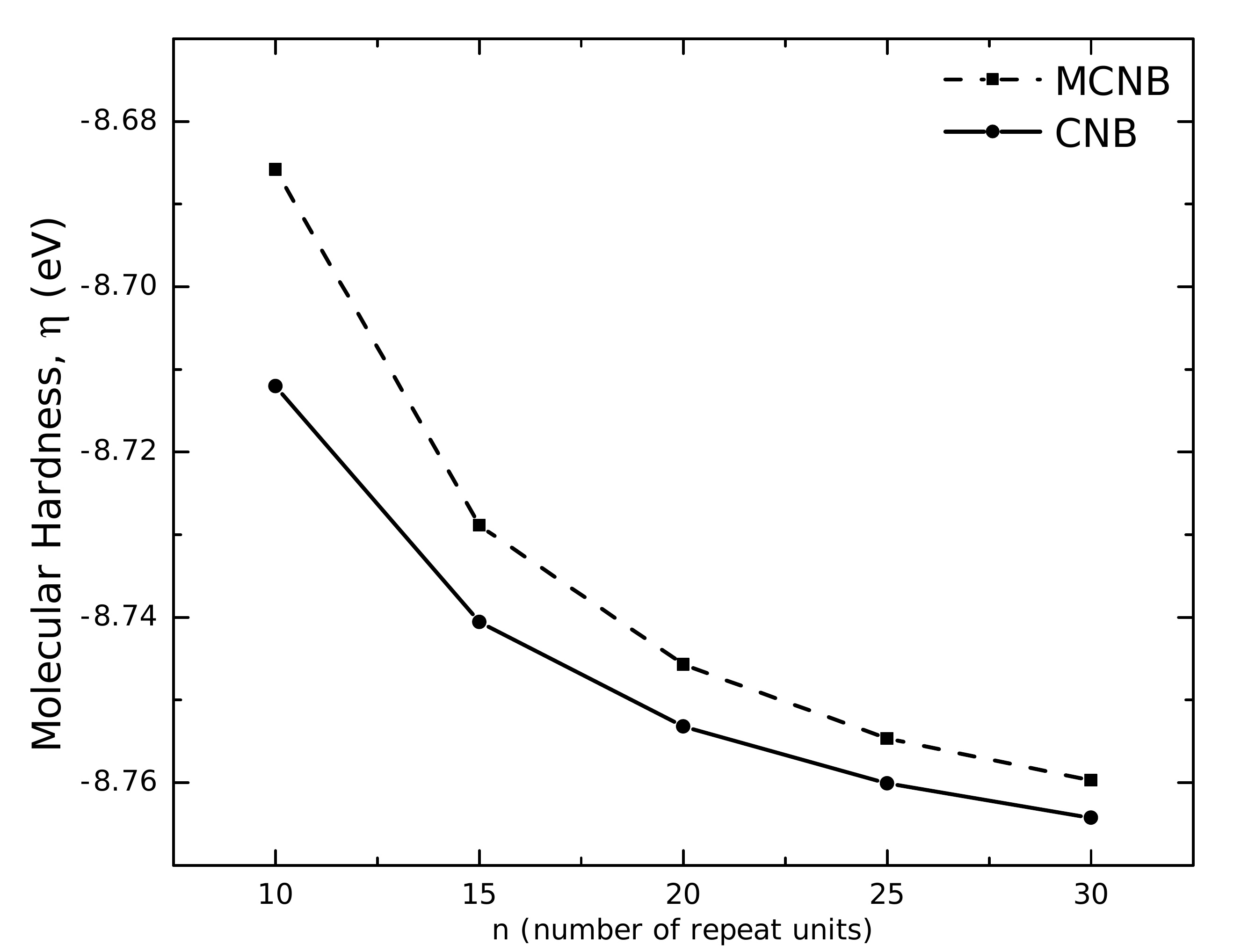}
\label{Fig:ChemicalPotential-C}}
\subfigure[][Molecular Hardness]{\includegraphics[width=\sizeA]{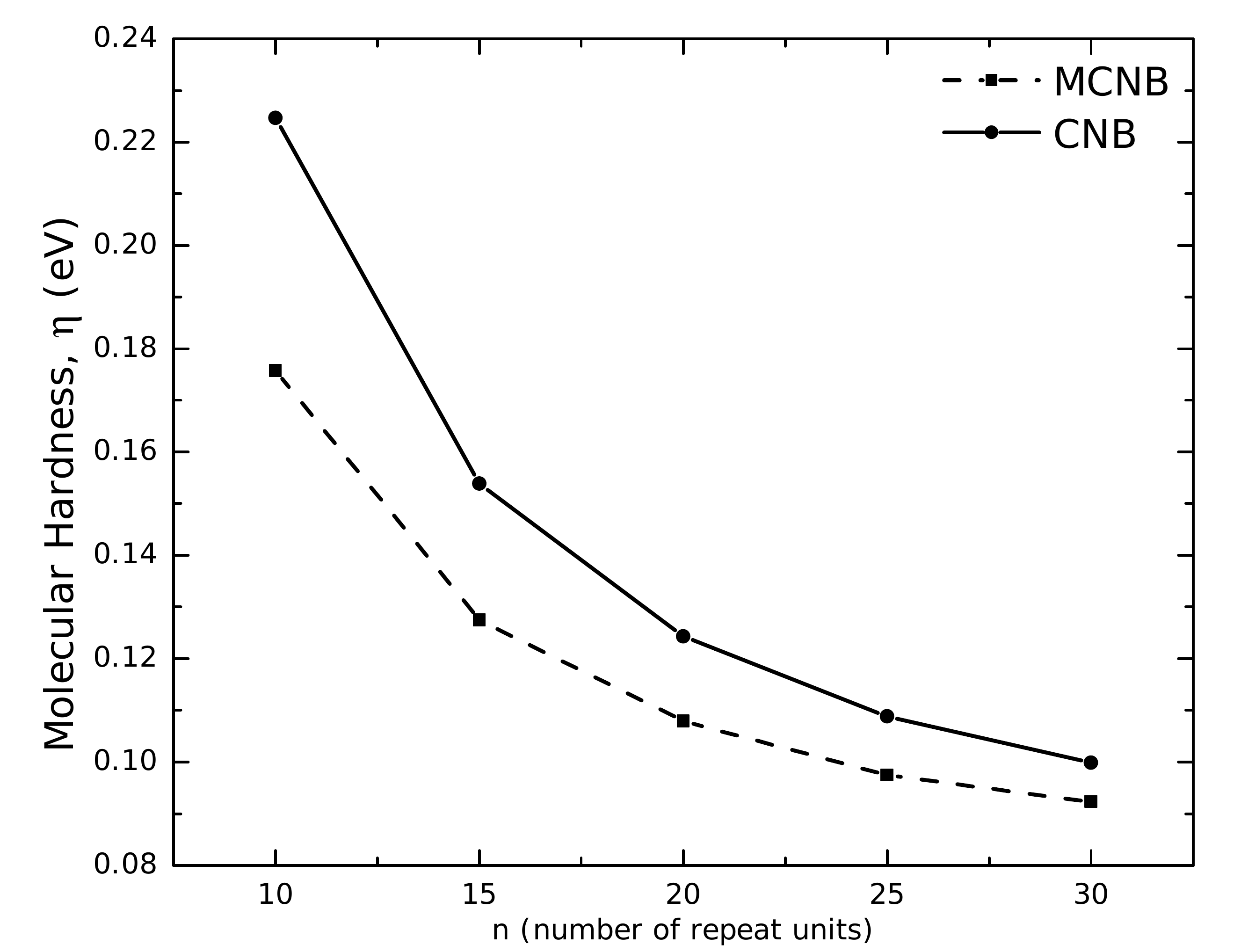}
\label{Fig:MolecularHardness-C}}
\subfigure[][Electrophilicity
Index]{\includegraphics[width=\sizeA]{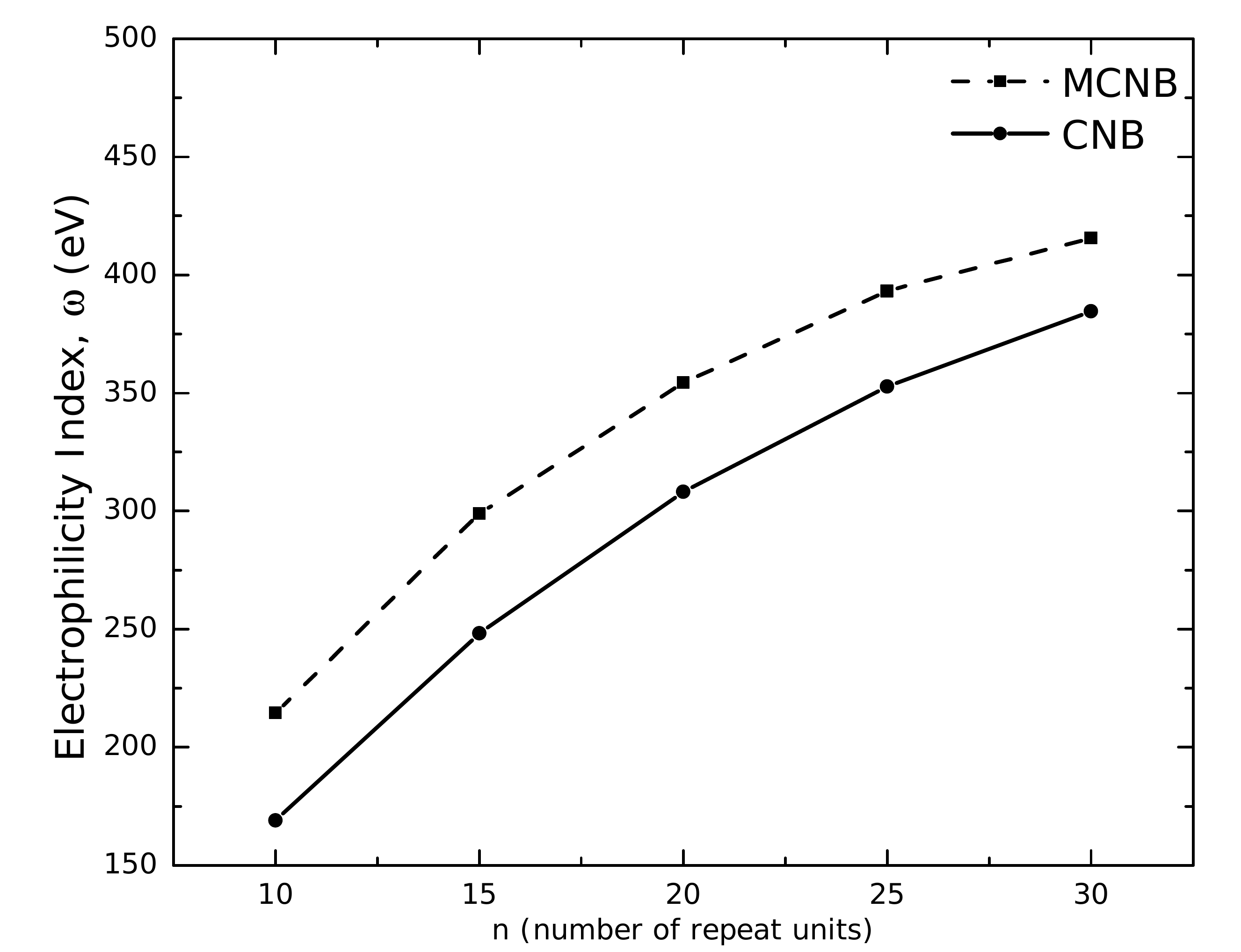}
\label{Fig:Electrophilicity-C}}
\caption{\label{Fig:ELECT_PROP-C} Calculated electronic properties for carbon
nanobelts.}
\end{figure}

Figure~\ref{Fig:ELECT_PROP-C} shows the electronic properties for the carbon
nanobelts. All the graphs shown a different behavior when compare to
boron--nitride nanobelts. In this case, the properties for both systems, CNB
and MCNB, behave with the same monotonic variation.

\renewcommand{\sizeA}{3.5cm}
\begin{figure}[tbph]
\centering
\begin{tabular}{ccc}
\subfigure[CNB\textsubscript{20}
structure]{\includegraphics[width=\sizeA]{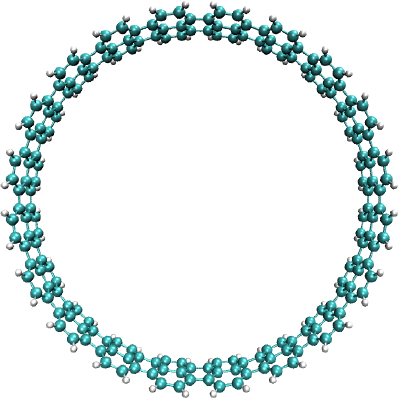}} &
\subfigure[CNB\textsubscript{20}
HOMO]{\includegraphics[width=\sizeA]{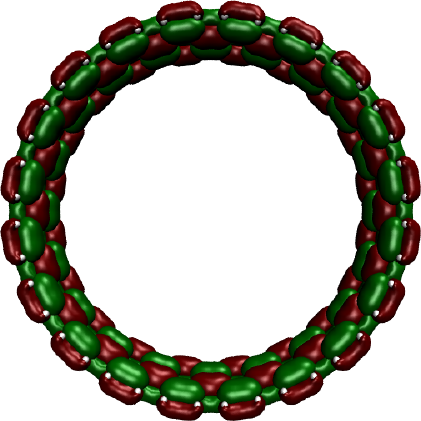}} &
\subfigure[CNB\textsubscript{20}
LUMO]{\includegraphics[width=\sizeA]{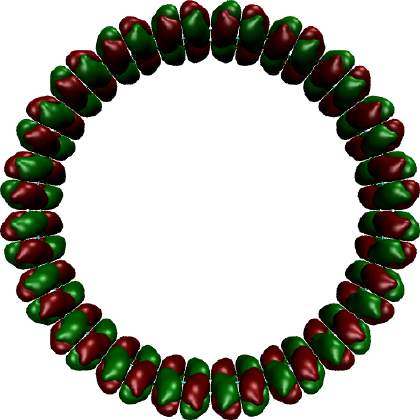}} \\

\subfigure[MCNB\textsubscript{20}
structure]{\includegraphics[width=\sizeA]{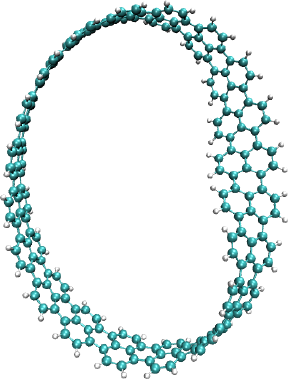}} &
\subfigure[MCNB\textsubscript{20}
HOMO]{\includegraphics[width=\sizeA]{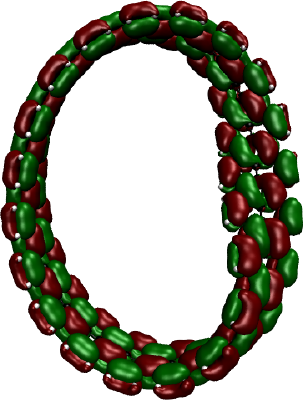}} &
\subfigure[MCNB\textsubscript{20}
LUMO]{\includegraphics[width=\sizeA]{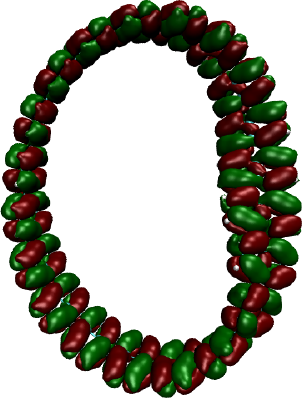}} \\

\end{tabular}
\caption{\label{Fig:OrbitalsCNB20} Structures, HOMO and LUMO surfaces for
CNB\textsubscript{20} and MCNB\textsubscript{20} systems. Red (green) color
represents negative (positive) values. Orbital surfaces rendered with with
isovalue equal to $0.001$ and with VMD~\cite{vmd} software.}
\end{figure}

The 3--D top view diagrams of the HOMO and LUMO surfaces for the
CNB\textsubscript{20} and MCNB\textsubscript{20} are shown in
figures~\ref{Fig:OrbitalsCNB20} (the 3--D
diagrams for all the
structures are shown in figures~\ref{FigS:OrbitalsCNB} and
\ref{FigS:OrbitalsMCNB} of Supplementary material). In this case, the torsion
on the nanobelt did not modify substantially either orbitals surfaces for any
system.

The values of $\Delta \varepsilon$, $\mu$, $\eta$ and $\omega$ for CNB and MCNB
systems, are similar among them having small variations with the increase of
the number of repeat units used to build the nanobelts. As the properties also
vary with the increase of repeat units, the change in n can tailor the
electronic properties of the carbon nanobelts.

The inhomogeneous distribution of the HOMO surfaces can be correlated with the
molecular hardness, $\eta$. As $\eta$ is related with how energetically
unfavorable is to redistribute the electrons in a molecule, higher values
implies a harder rearrangement of them. Comparing the values of $\eta$
for the MBNNB (\ref{Fig:MolecularHardness-BN}) and MCNB
(\ref{Fig:MolecularHardness-C}) systems, we can see that $\eta$ is almost eight
times bigger for boron--nitride nanobelts than for carbon based nanobelts. A
lower value of $\eta$ imply that the electrons can redistribute easily through
the whole structure resulting in a more homogeneous HOMO surface.

\section{Conclusions}
In this work we studied the structural and electronic properties of
boron--nitride and carbon based M\"obius nanobelts and compare their properties
with simple nanobelts. For all systems, the main peaks in the infrared
calculated spectra are in accordance with the experimental one, indicating that
the theoretical methodology used here is suitable to determine other properties.

The electronic properties shown differences between both boron--nitride
nanobelts. Whereas the LUMO energy for both systems, BNNB and MBNNB, have
similar monotony, the other properties shown opposite behavior (different
monotony). All the properties for the carbon based nanobelts have the same
monotonic behavior for all the properties. Finally, the inhomogeneous
distribution of the HOMO surface for the MBNNB was related with the high values
of the molecular hardness. In all cases, the properties vary
with the increase of the number of repeat units indicating that it is possible
to choose the desired values changing the size and type of the systems.

\section*{Acknowledgements}
We would like to acknowledge financial support from the Brazilian agencies
CNPq, CAPES and FAPEMIG. Part of the results presented here were developed
with the help of a CENAPAD-SP (Centro Nacional de Processamento de Alto
Desempenho em S\~ao Paulo) grant UNICAMP/FINEP--MCT, CENAPAD--UFC (Centro
Nacional de Processamento de Alto Desempenho, at Universidade Federal do
Cear\'a), and Digital Research Alliance of
Canada (via  project bmh-491-09 belonging to Dr. Nike Dattani), for the
computational support.


\begin{thebibliography}{10}
\expandafter\ifx\csname url\endcsname\relax
  \def\url#1{\texttt{#1}}\fi
\expandafter\ifx\csname urlprefix\endcsname\relax\def\urlprefix{URL }\fi
\expandafter\ifx\csname href\endcsname\relax
  \def\href#1#2{#2} \def\path#1{#1}\fi

\bibitem{Kasalkova-Nanomaterials-11-2368-2021}
N.~S. Kas{\'{a}}lkov{\'{a}}, P.~Slepi{\v{c}}ka, V.~{\v{S}}vor{\v{c}}{\'{\i}}k,
  Carbon nanostructures, nanolayers, and their composites, Nanomaterials 11
  (2021) 2368 (2021).
\newblock \href {https://doi.org/10.3390/nano11092368}
  {\path{doi:10.3390/nano11092368}}.

\bibitem{Pedrosa-MaterialsResearch-23-20190493-2020}
M.~C.~G. Pedrosa, J.~C.~D. Filho, L.~R. {de Menezes}, E.~O. {da Silva},
  Chemical surface modification and characterization of carbon nanostructures
  without shape damage, Materials Research 23 (2020) e20190493 (2020).
\newblock \href {https://doi.org/10.1590/1980-5373-MR-2019-0493}
  {\path{doi:10.1590/1980-5373-MR-2019-0493}}.

\bibitem{Jirimali-Materials-15-3969-2022}
H.~Jirimali, J.~Singh, R.~Boddula, J.~Lee, V.~Singh, {Nano-structured carbon:
  Its synthesis from renewable agricultural sources and important
  applications}, Materials 15 (2022) 3969 (2022).
\newblock \href {https://doi.org/10.3390/ma15113969}
  {\path{doi:10.3390/ma15113969}}.

\bibitem{Segawa-Science-365-272-2019}
Y.~Segawa, M.~Kuwayama, Y.~Hijikata, M.~Fushimi, T.~Nishihara, J.~Pirillo,
  J.~Shirasaki, N.~Kubota, K.~Itami, {Topological molecular nanocarbons:
  All-benzene catenane and trefoil knot}, Science 365 (2019) 272--276 (2019).
\newblock \href {https://doi.org/10.1126/science.aav5021}
  {\path{doi:10.1126/science.aav5021}}.

\bibitem{Cheung-Chem.Eur.J.-26-14791-2020}
K.~Y. Cheung, Y.~Segawa, K.~Itami, Synthetic strategies of carbon nanobelts and
  related belt-shaped polycyclic aromatic hydrocarbons, Chem. Eur. J. 26 (2020)
  14791--14801 (2020).
\newblock \href {https://doi.org/10.1002/chem.202002316}
  {\path{doi:10.1002/chem.202002316}}.

\bibitem{Panwar-Chem.Rev.-119-9559-2019}
N.~Panwar, A.~M. Soehartono, K.~K. Chan, S.~Zeng, G.~Xu, J.~Qu, P.~Coquet,
  K.~Yong, X.~Chen, {Nanocarbons for biology and medicine: Sensing, imaging,
  and drug delivery}, Chem. Rev. 119~(1) (2019) 9559--9656 (2019).
\newblock \href {https://doi.org/10.1021/acs.chemrev.9b00099}
  {\path{doi:10.1021/acs.chemrev.9b00099}}.

\bibitem{Saba_2019}
N.~Saba, M.~Jawaid, H.~Fouad, O.~Y. Alothman, Nanocarbon: Preparation,
  properties, and applications, Elsevier, 2019 (2019).
\newblock \href {https://doi.org/10.1016/B978-0-08-102509-3.00009-2}
  {\path{doi:10.1016/B978-0-08-102509-3.00009-2}}.

\bibitem{Shearer-275-2020}
C.~J. Shearer, L.~Yu, J.~G. Shapter, Synthesis and applications of nanocarbons,
  Wiley, 2020, Ch. Optoelectronic properties of nanocarbons and nanocarbon
  films, pp. 275--294 (2020).
\newblock \href {https://doi.org/10.1002/9781119429418.ch9}
  {\path{doi:10.1002/9781119429418.ch9}}.

\bibitem{Itami-NanoLett.-20-4718-2020}
K.~Itami, T.~Maekawa, {Molecular nanocarbon science: Present and future}, Nano
  Lett. 20 (2020) 4718--4720 (2020).
\newblock \href {https://doi.org/10.1021/acs.nanolett.0c02143}
  {\path{doi:10.1021/acs.nanolett.0c02143}}.

\bibitem{Guo-Nat.Chem.-13-402-2021}
Q.~Guo, Y.~Qiu, M.~Wang, J.~F. Stoddart, Aromatic hydrocarbon belts, Nat. Chem.
  13 (2021) 402--419 (2021).
\newblock \href {https://doi.org/10.1038/s41557-021-00671-9}
  {\path{doi:10.1038/s41557-021-00671-9}}.

\bibitem{Yang-Chem.Rev.-120-2693-2020}
F.~Yang, M.~Wang, D.~Zhang, J.~Yang, M.~Zheng, Y.~Li, {Chirality pure carbon
  nanotubes: Growth, sorting, and characterization}, Chem. Rev. 120 (2020)
  2693--2758 (2020).
\newblock \href {https://doi.org/10.1021/acs.chemrev.9b00835}
  {\path{doi:10.1021/acs.chemrev.9b00835}}.

\bibitem{Povie-Science-356-172-2017}
G.~Povie, Y.~Segawa, T.~Nishihara, Y.~Miyauchi, K.~Itami, Synthesis of a carbon
  nanobelt, Science 356 (2017) 172--175 (2017).
\newblock \href {https://doi.org/10.1126/science.aam8158}
  {\path{doi:10.1126/science.aam8158}}.

\bibitem{Xia-Angew.Chem.-133-10399-2021}
Z.~Xia, S.~H. Pun, H.~Chen, Q.~Miao, Synthesis of zigzag carbon nanobelts
  through scholl reactions, Angew. Chem. 133 (2021) 10399--10406 (2021).
\newblock \href {https://doi.org/10.1002/ange.202100343}
  {\path{doi:10.1002/ange.202100343}}.

\bibitem{Price-NatureSynthesis-1-502-2022}
T.~W. Price, R.~Jasti, Carbon nanobelts do the twist, Nature Synthesis 1 (2022)
  502--503 (2022).
\newblock \href {https://doi.org/10.1038/s44160-022-00083-8}
  {\path{doi:10.1038/s44160-022-00083-8}}.

\bibitem{Lu-Chem-2-619-2017}
X.~Lu, J.~Wu, After 60 years of efforts: The chemical synthesis of a carbon
  nanobelt, Chem 2 (2017) 619--620 (2017).
\newblock \href {https://doi.org/10.1016/j.chempr.2017.04.012}
  {\path{doi:10.1016/j.chempr.2017.04.012}}.

\bibitem{Chen-J.Phys.Org.Chem.-33--2020}
H.~Chen, Q.~Miao, Recent advances and attempts in synthesis of conjugated
  nanobelts, J. Phys. Org. Chem. 33 (2020).
\newblock \href {https://doi.org/10.1002/poc.4145}
  {\path{doi:10.1002/poc.4145}}.

\bibitem{Segawa-NatureSynthesis-1-535-2022}
Y.~Segawa, T.~Watanabe, K.~Yamanoue, M.~Kuwayama, K.~Watanabe, J.~Pirillo,
  Y.~Hijikata, K.~Itami, Synthesis of a {M\"bius} carbon nanobelt, Nature
  Synthesis 1 (2022) 535--541 (2022).
\newblock \href {https://doi.org/10.1038/s44160-022-00075-8}
  {\path{doi:10.1038/s44160-022-00075-8}}.

\bibitem{Li-Acc.Mater.Res.-2-681-2021}
Y.~Li, H.~Kono, T.~Maekawa, Y.~Segawa, A.~Yagi, K.~Itami, Chemical synthesis of
  carbon nanorings and nanobelts, Acc. Mater. Res. 2 (2021) 681--691 (2021).
\newblock \href {https://doi.org/10.1021/accountsmr.1c00105}
  {\path{doi:10.1021/accountsmr.1c00105}}.

\bibitem{Nishigaki-J.Am.Chem.Soc.-141-14955-2019}
S.~Nishigaki, Y.~Shibata, A.~Nakajima, H.~Okajima, Y.~Masumoto, T.~Osawa,
  A.~Muranaka, H.~Sugiyama, A.~Horikawa, H.~Uekusa, H.~Koshino, M.~Uchiyama,
  A.~Sakamoto, K.~Tanaka, Synthesis of belt- and {M\"obius}-shaped
  cycloparaphenylenes by rhodium-catalyzed alkyne cyclotrimerization, J. Am.
  Chem. Soc. 141 (2019) 14955--14960 (2019).
\newblock \href {https://doi.org/10.1021/jacs.9b06197}
  {\path{doi:10.1021/jacs.9b06197}}.

\bibitem{Wang-Angew.Chem.Int.Ed.-60-18443-2021}
S.~Wang, J.~Yuan, J.~Xie, Z.~Lu, L.~Jiang, Y.~Mu, Y.~Huo, Y.~Tsuchido, K.~Zhu,
  {Sulphur-embedded hydrocarbon belts: Synthesis, structure and redox chemistry
  of cyclothianthrenes}, Angew. Chem. Int. Ed. 60 (2021) 18443--18447 (2021).
\newblock \href {https://doi.org/10.1002/anie.202104054}
  {\path{doi:10.1002/anie.202104054}}.

\bibitem{Ajami-Nature-426-819-2003}
D.~Ajami, O.~Oeckler, A.~Simon, R.~Herges, Synthesis of a {M\"obius} aromatic
  hydrocarbon, Nature 426 (2003) 819--821 (2003).
\newblock \href {https://doi.org/10.1038/nature02224}
  {\path{doi:10.1038/nature02224}}.

\bibitem{VNL}
{Virtual NanoLab - Atomistix ToolKit. QuantumWise. v2017.1} (2017).

\bibitem{xTB_1}
C.~Bannwarth, E.~Caldeweyher, S.~Ehlert, A.~Hansen, P.~Pracht, J.~Seibert,
  S.~Spicher, S.~Grimme, Extended tight-binding quantum chemistry methods,
  WIREs Comput. Mol. Sci. 11 (2020) e1493 (2020).
\newblock \href {https://doi.org/10.1002/wcms.1493}
  {\path{doi:10.1002/wcms.1493}}.

\bibitem{xTB_3}
C.~Bannwarth, S.~Ehlert, S.~Grimme, {GFN2--xTB}--an accurate and broadly
  parametrized self-consistent tight-binding quantum chemical method with
  multipole electrostatics and density-dependent dispersion contributions, J.
  Chem. Theory Comput. 15 (2019) 1652--1671 (2019).
\newblock \href {https://doi.org/10.1021/acs.jctc.8b01176}
  {\path{doi:10.1021/acs.jctc.8b01176}}.

\bibitem{koopmans-Physica-1-104-1933}
T.~Koopmans, {\"Uber} die zuordnung von wellenfunktionen und eigenwerten zu den
  einzelnen elektronen eines atoms, Physica 1 (1933) 104--113 (1933).

\bibitem{luo-J.Phys.Chem.A-110-12005-2006}
J.~Luo, Z.~Q. Xue, W.~M. Liu, J.~L. Wu, Z.~Q. Yang, Koopmans' theorem for large
  molecular systems within density functional theory, J. Phys. Chem. A 110
  (2006) 12005--12009 (2006).

\bibitem{salzner-J.Chem.Phys.-131-231101-2009}
U.~Salzner, R.~Baer, Koopmans' springs to life, J. Chem. Phys. 131 (2009)
  231101 (2009).

\bibitem{tsuneda-J.Chem.Phys.-133-174101-2010}
T.~Tsuneda, J.~W. Song, S.~Suzuki, K.~Hirao, On {Koopmans'} theorem in density
  functional theory, J. Chem. Phys. 133 (2010) 174101 (2010).

\bibitem{janak-Phys.Rev.-1978-7165-18}
J.~F. Janak, Proof that ${{\partial E} \mathord{\left/{\vphantom {{\partial E}
  {\partial {n_i} = {\varepsilon _i}}}} \right.\kern-\nulldelimiterspace}
  {\partial {n_i} = {\varepsilon _i}}}$ in density-functional theory, Phys.
  Rev. 1978 (18) 7165--7168 (18).

\bibitem{zhan-J.Phys.Chem.A-107-4184-2003}
C.~G. Zhan, J.~A. Nichols, D.~A. Dixon, Ionization potential, electron
  affinity, electronegativity, hardness, and electron excitation energy:
  {M}olecular properties from density functional theory orbital energies, J.
  Phys. Chem. A 107 (2003) 4184--4195 (2003).

\bibitem{parr-J.Am.Chem.Soc.-121-1922-1999}
R.~G. Parr, L.~V. Szentp\'aly, S.~Liu, Electrophilicity index, J. Am. Chem.
  Soc. 121 (1999) 1922--1924 (1999).

\bibitem{vmd}
W.~Humphrey, A.~Dalke, K.~Schulten, {VMD}: Visual molecular dynamics, Journal
  of Molecular Graphics 14 (1996) 33--38 (fe 1996).
\newblock \href {https://doi.org/10.1016/0263-7855(96)00018-5}
  {\path{doi:10.1016/0263-7855(96)00018-5}}.

\bibitem{Wirtz-Phys.Rev.B-68-45425-2003}
L.~Wirtz, A.~Rubio, R.~A. {de la Concha}, A.~Loiseau, \emph{Ab initio}
  calculations of the lattice dynamics of boron nitride nanotubes, Phys. Rev. B
  68 (2003) 045425 (2003).
\newblock \href {https://doi.org/10.1103/PhysRevB.68.045425}
  {\path{doi:10.1103/PhysRevB.68.045425}}.

\bibitem{Singh-Sci.Rep.-6-35535-2016}
B.~Singh, G.~Kaur, P.~Singh, K.~Singh, B.~Kumar, A.Vij, M.~Kumar, R.~Bala,
  R.~Meena, A.~Singh, A.~Thakur, A.~Kumar, Nanostructured boron nitride with
  high water dispersibility for boron neutron capture therapy, Sci. Rep. 6
  (2016) 35535 (2016).
\newblock \href {https://doi.org/10.1038/srep35535}
  {\path{doi:10.1038/srep35535}}.

\bibitem{Harrison-NanoscaleAdv.-1-1693-2019}
H.~Harrison, J.~T. Lamb, K.~S. Nowlin, A.~J. Guenthner, K.~B. Ghiassi, A.~D.
  Kelkar, J.~R. Alston, Quantification of hexagonal boron nitride impurities in
  boron nitride nanotubes \emph{via} {FTIR} spectroscopy, Nanoscale Adv. 1
  (2019) 1693--1701 (2019).
\newblock \href {https://doi.org/10.1039/c8na00251g}
  {\path{doi:10.1039/c8na00251g}}.

\bibitem{Zhang-Catalysts-10-596-2020}
H.~Zhang, B.-G. Shin, D.-E. Lee, K.-B. Yoon, Preparation of {PP/2D}--nanosheet
  composites using {$MoS_2$/$MgCl_2$}-- and {$BN$/$MgCl_2$}--bisupported
  {Ziegler--Natta} catalysts, Catalysts 10 (2020) 596 (2020).
\newblock \href {https://doi.org/10.3390/catal10060596}
  {\path{doi:10.3390/catal10060596}}.

\bibitem{Nandiyanto-Indones.J.Sci.Technol.-4-97-2019}
A.~B.~D. Nandiyanto, R.~Oktiani, R.~Ragadhita, How to read and interpret {FTIR}
  spectroscope of organic material, Indones. J. Sci. Technol. 4 (2019) 97--118
  (2019).
\newblock \href {https://doi.org/10.17509/ijost.v4i1.15806}
  {\path{doi:10.17509/ijost.v4i1.15806}}.

\bibitem{zhang-J.Phys.Chem.A-111-1554-2007}
G.~Zhang, C.~Musgrave, Comparison of {DFT} methods for molecular orbital
  eigenvalue calculations, J. Phys. Chem. A 111 (2007) 1554--1561 (2007).

\bibitem{parr-J.Am.Chem.Soc.-105-7512-1983}
R.~G. Parr, R.~G. Pearson, Absolute hardness: companion parameter to absolute
  electronegativity, J. Am. Chem. Soc. 105 (1983) 7512--7516 (1983).

\end{thebibliography}

\newpage
\begin{center}
\textbf{\large Supplementary material: Electronic and structural properties of
M\"obius boron--nitride and carbon nanobelts}
\end{center}

\setcounter{section}{0}
\setcounter{equation}{0} \setcounter{figure}{0} \setcounter{table}{0}
\setcounter{page}{1} \makeatletter
\renewcommand{\thesection}{S\arabic{section}}
\renewcommand{\theequation}{S\arabic{equation}}
\renewcommand{\thefigure}{S\arabic{figure}}
\renewcommand{\bibnumfmt}[1]{[S#1]}
\renewcommand{\citenumfont}[1]{S#1}

\newcommand{\sizeC}{2.5cm}
\begin{figure}[tbph]
\centering
\begin{tabular}{ccc}
BNNB\textsubscript{x} & HOMO & LUMO \\
\subfigure{\includegraphics[width=\sizeC]{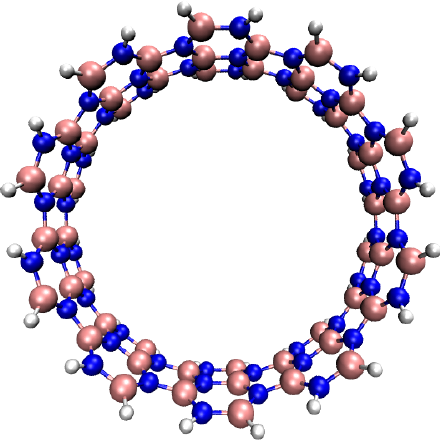}}
 &
\subfigure{\includegraphics[width=\sizeC]{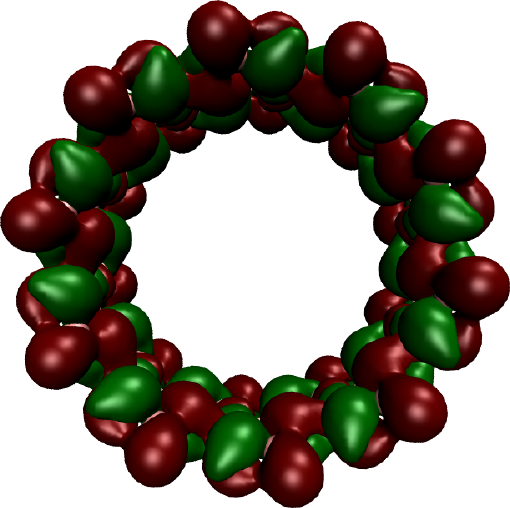}} &
\subfigure{\includegraphics[width=\sizeC]{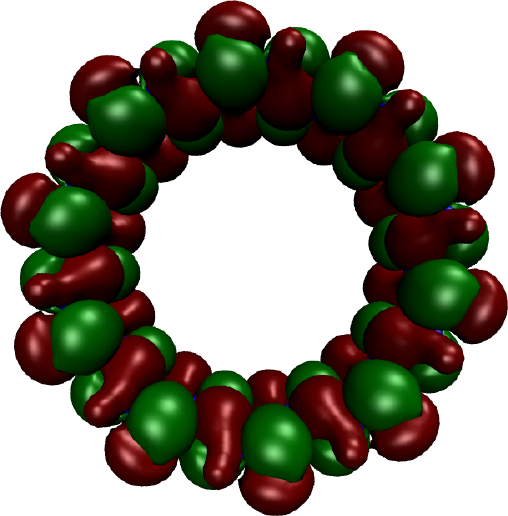}} \\

\subfigure{\includegraphics[width=\sizeC]{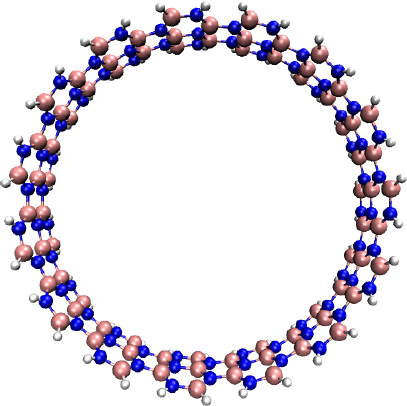}}
 &
\subfigure{\includegraphics[width=\sizeC]{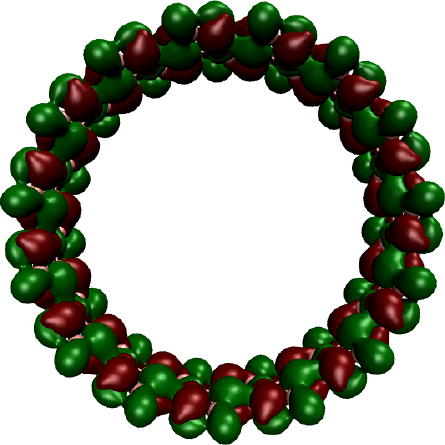}} &
\subfigure{\includegraphics[width=\sizeC]{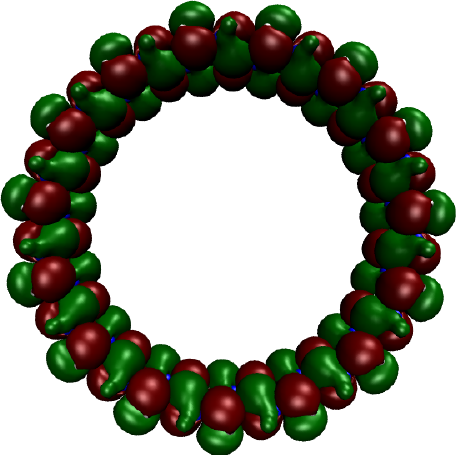}} \\

\subfigure{\includegraphics[width=\sizeC]{fig_STRUCT_BNNB20x.pdf}}
 &
\subfigure{\includegraphics[width=\sizeC]{fig_HOMO_BNNB20x.pdf}} &
\subfigure{\includegraphics[width=\sizeC]{fig_LUMO_BNNB20x.pdf}} \\

\subfigure{\includegraphics[width=\sizeC]{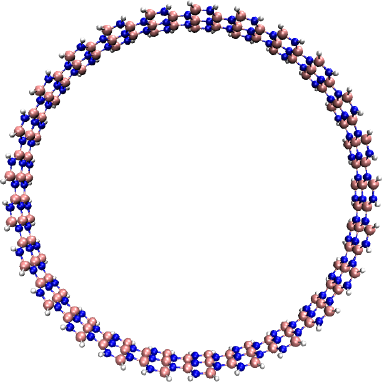}}
 &
\subfigure{\includegraphics[width=\sizeC]{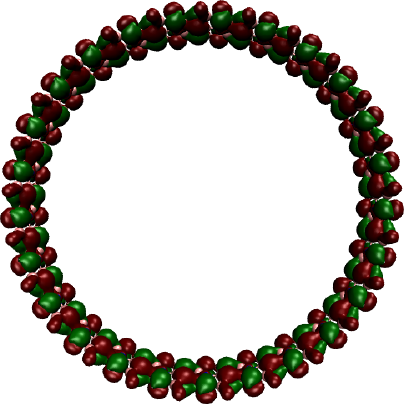}} &
\subfigure{\includegraphics[width=\sizeC]{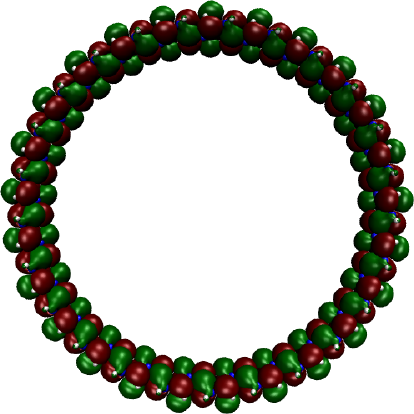}} \\

\subfigure{\includegraphics[width=\sizeC]{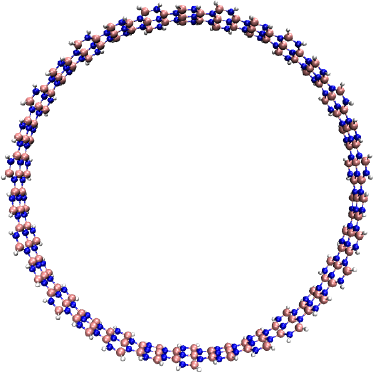}}
 &
\subfigure{\includegraphics[width=\sizeC]{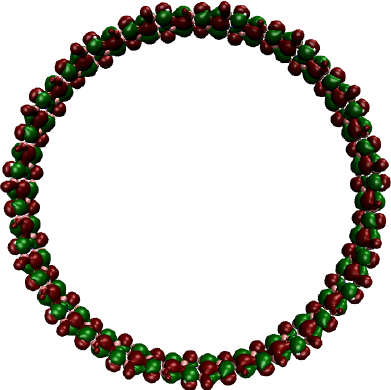}} &
\subfigure{\includegraphics[width=\sizeC]{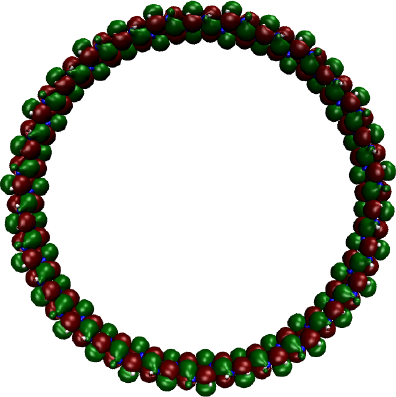}} \\

\end{tabular}
\caption{\label{FigS:OrbitalsBNNB} Frontier orbitals (HOMO and LUMO) for all
boron--nitride nanobelts. Top to bottom: number of repetitions from 10 to 30.}
\end{figure}
\restoregeometry

\newpage
\begin{figure}[tbph]
\centering
\begin{tabular}{ccc}
MBNNB\textsubscript{x} & HOMO & LUMO \\
\subfigure{\includegraphics[width=\sizeC]{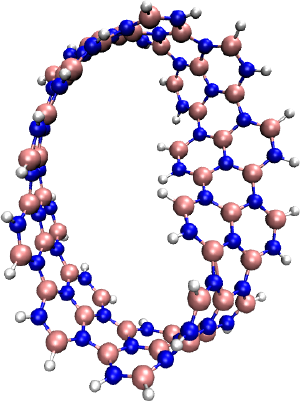}} &
\subfigure{\includegraphics[width=\sizeC]{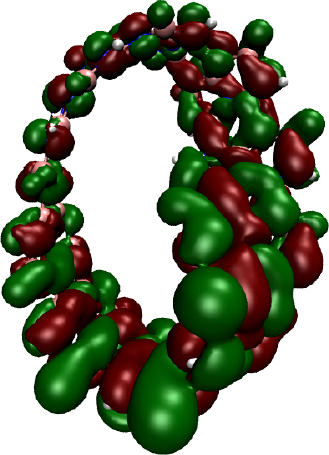}} &
\subfigure{\includegraphics[width=\sizeC]{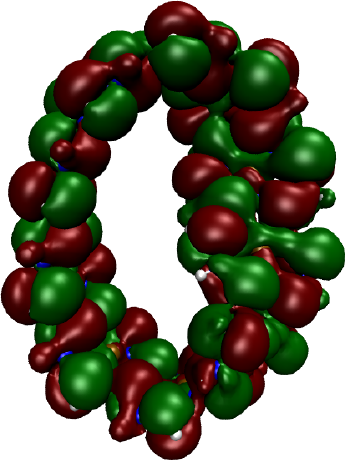}} \\

\subfigure{\includegraphics[width=\sizeC]{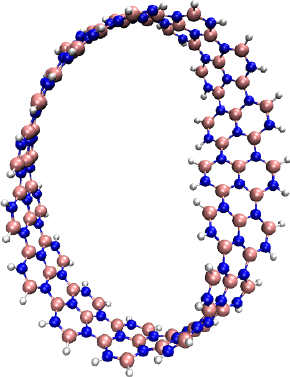}} &
\subfigure{\includegraphics[width=\sizeC]{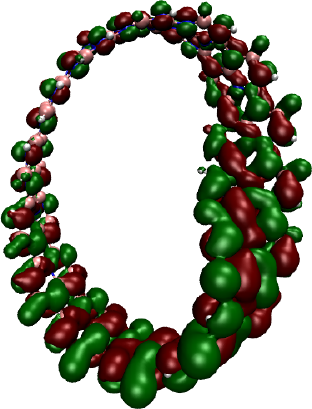}} &
\subfigure{\includegraphics[width=\sizeC]{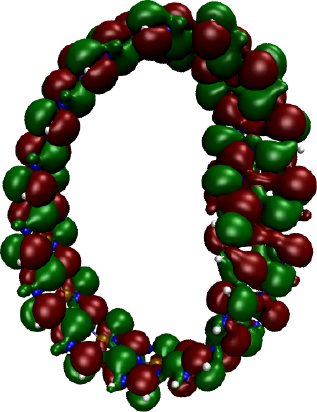}} \\

\subfigure{\includegraphics[width=\sizeC]{fig_STRUCT_MBNNB20x.pdf}} &
\subfigure{\includegraphics[width=\sizeC]{fig_HOMO_MBNNB20x.pdf}} &
\subfigure{\includegraphics[width=\sizeC]{fig_LUMO_MBNNB20x.pdf}} \\

\subfigure{\includegraphics[width=\sizeC]{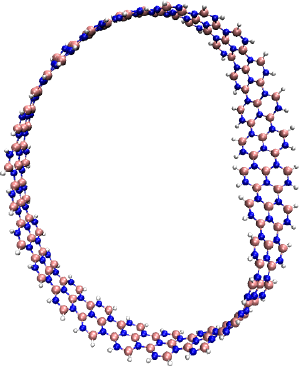}} &
\subfigure{\includegraphics[width=\sizeC]{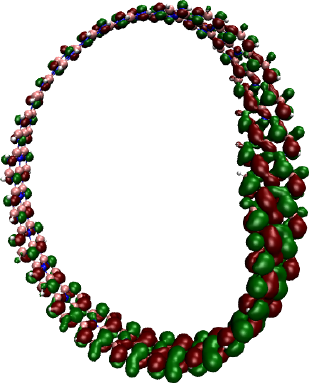}} &
\subfigure{\includegraphics[width=\sizeC]{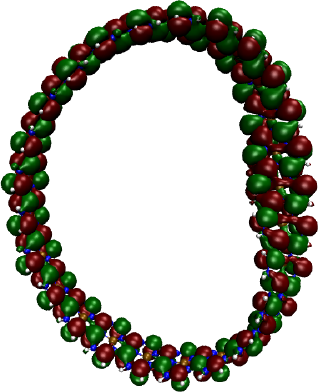}} \\

\subfigure{\includegraphics[width=\sizeC]{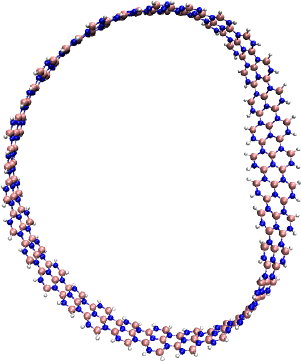}} &
\subfigure{\includegraphics[width=\sizeC]{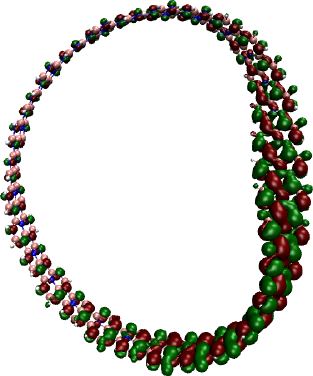}} &
\subfigure{\includegraphics[width=\sizeC]{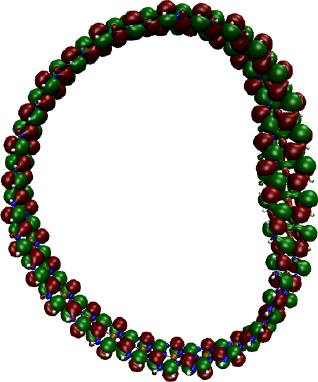}} \\

\end{tabular}
\caption{\label{FigS:OrbitalsMBNNB} Frontier orbitals (HOMO and LUMO) for all
M\"obius boron--nitride nanobelts. Top to bottom: number of repetitions from 10
to 30.}
\end{figure}
\restoregeometry

\newpage
\begin{figure}[tbph]
\centering
\begin{tabular}{ccc}
CNB\textsubscript{x} & HOMO & LUMO \\
\subfigure{\includegraphics[width=\sizeC]{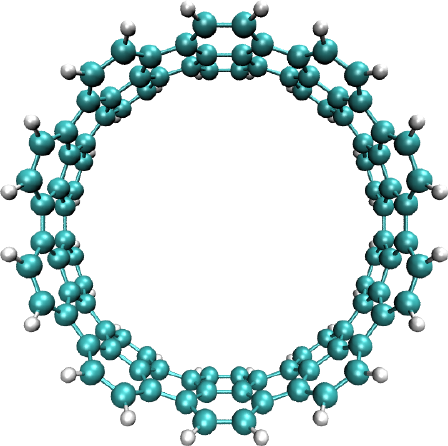}} &
\subfigure{\includegraphics[width=\sizeC]{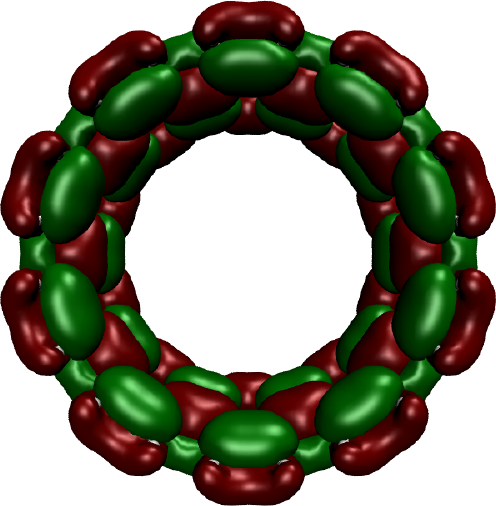}} &
\subfigure{\includegraphics[width=\sizeC]{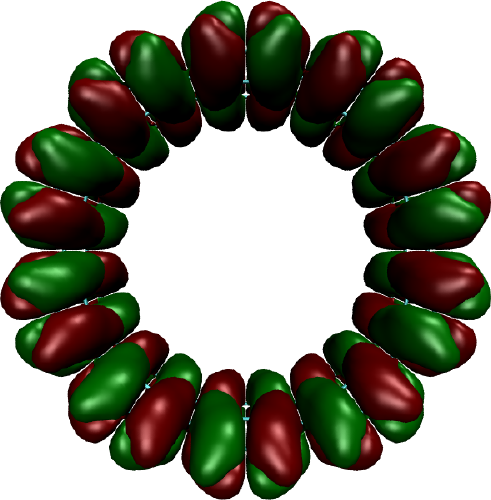}} \\

\subfigure{\includegraphics[width=\sizeC]{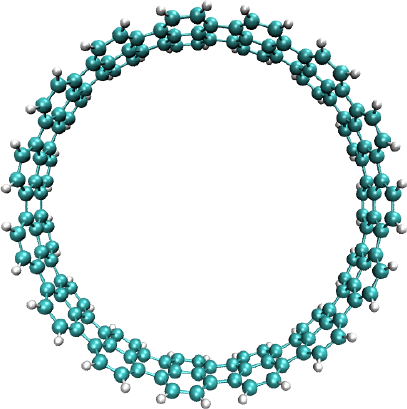}} &
\subfigure{\includegraphics[width=\sizeC]{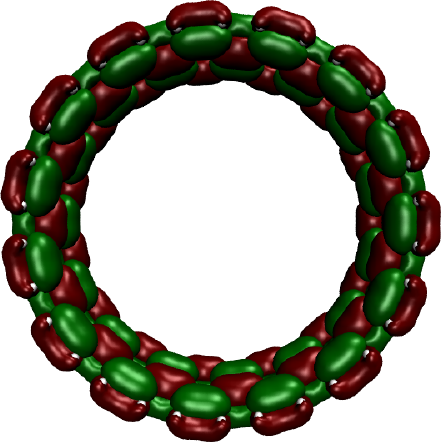}} &
\subfigure{\includegraphics[width=\sizeC]{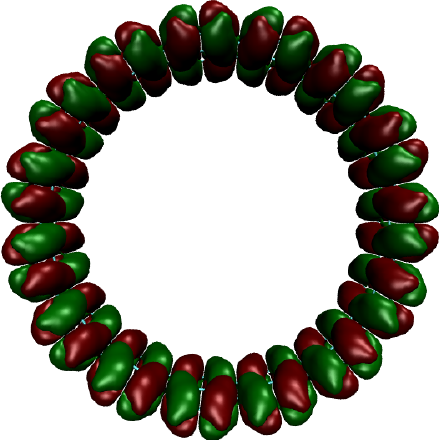}} \\

\subfigure{\includegraphics[width=\sizeC]{fig_STRUCT_CNB20x.pdf}} &
\subfigure{\includegraphics[width=\sizeC]{fig_HOMO_CNB20x.pdf}} &
\subfigure{\includegraphics[width=\sizeC]{fig_LUMO_CNB20x.pdf}} \\

\subfigure{\includegraphics[width=\sizeC]{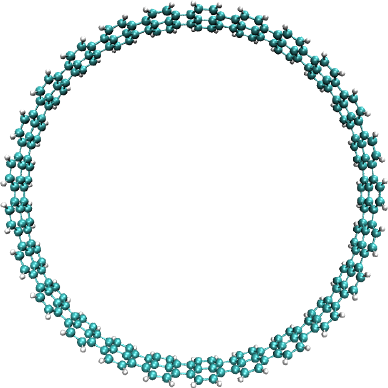}} &
\subfigure{\includegraphics[width=\sizeC]{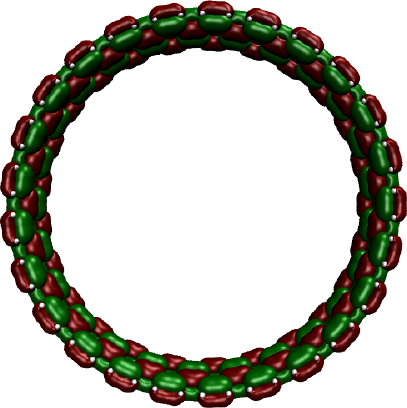}} &
\subfigure{\includegraphics[width=\sizeC]{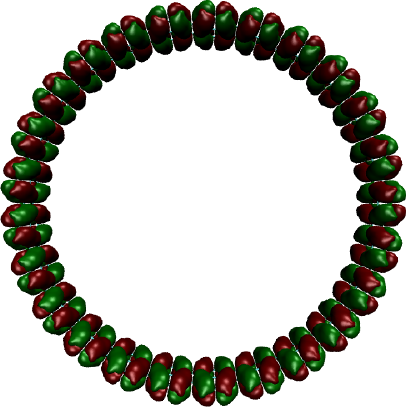}} \\

\subfigure{\includegraphics[width=\sizeC]{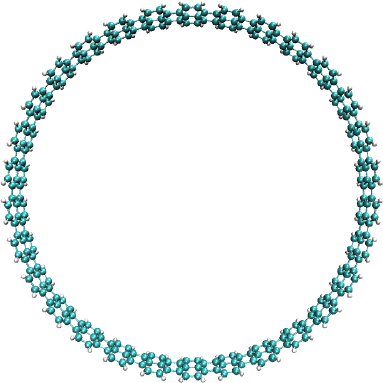}} &
\subfigure{\includegraphics[width=\sizeC]{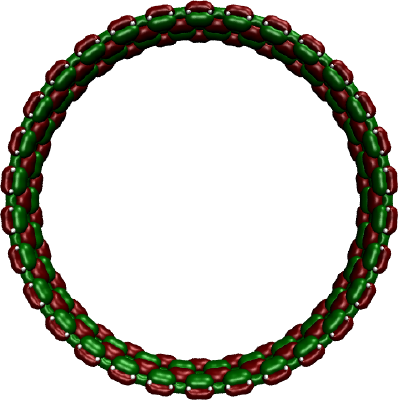}} &
\subfigure{\includegraphics[width=\sizeC]{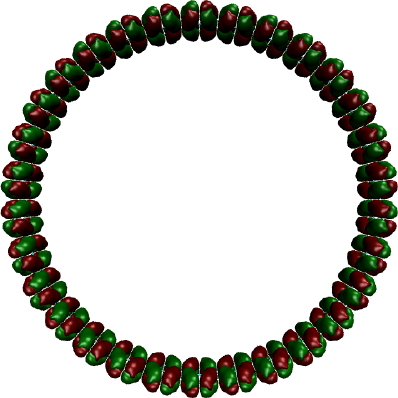}} \\

\end{tabular}
\caption{\label{FigS:OrbitalsCNB} Frontier orbitals (HOMO and LUMO) for all
carbon nanobelts. Top to bottom: number of repetitions from 10 to 30.}
\end{figure}
\restoregeometry

\newpage
\begin{figure}[tbph]
\centering
\begin{tabular}{ccc}
MCNB\textsubscript{x} & HOMO & LUMO \\
\subfigure{\includegraphics[width=\sizeC]{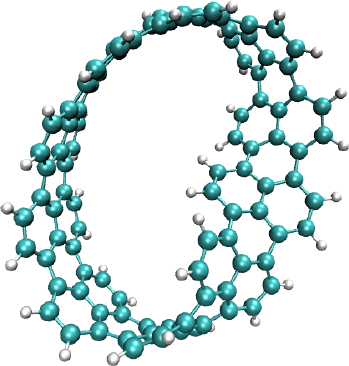}} &
\subfigure{\includegraphics[width=\sizeC]{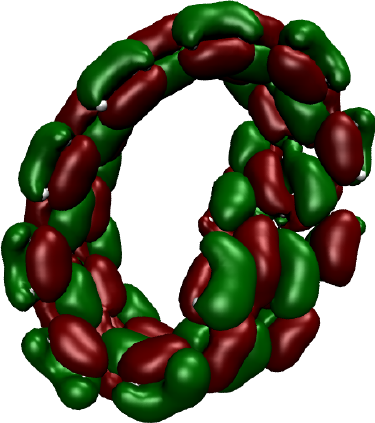}} &
\subfigure{\includegraphics[width=\sizeC]{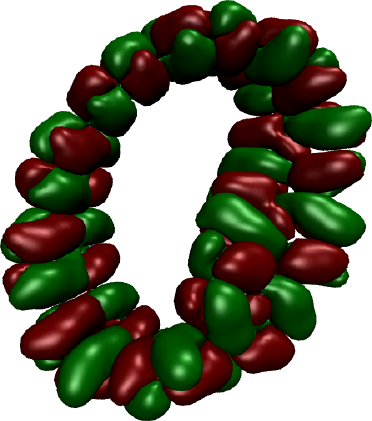}} \\

\subfigure{\includegraphics[width=\sizeC]{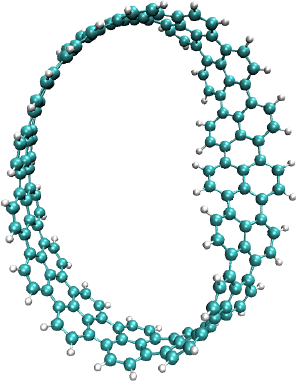}} &
\subfigure{\includegraphics[width=\sizeC]{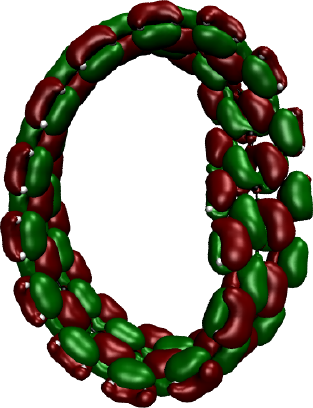}} &
\subfigure{\includegraphics[width=\sizeC]{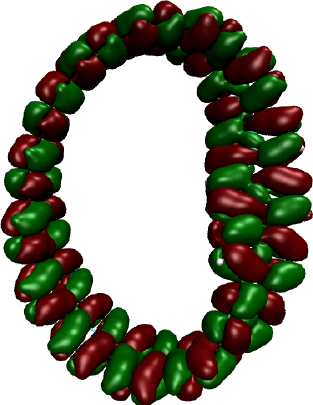}} \\

\subfigure{\includegraphics[width=\sizeC]{fig_STRUCT_MCNB20x.pdf}} &
\subfigure{\includegraphics[width=\sizeC]{fig_HOMO_MCNB20x.pdf}} &
\subfigure{\includegraphics[width=\sizeC]{fig_LUMO_MCNB20x.pdf}} \\

\subfigure{\includegraphics[width=\sizeC]{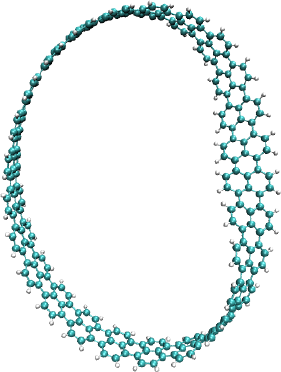}} &
\subfigure{\includegraphics[width=\sizeC]{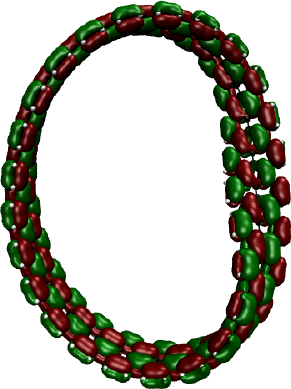}} &
\subfigure{\includegraphics[width=\sizeC]{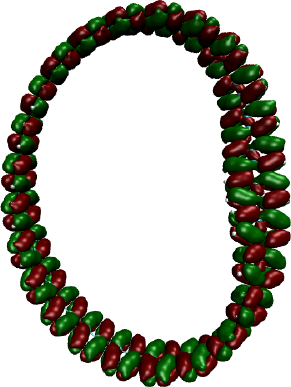}} \\

\subfigure{\includegraphics[width=\sizeC]{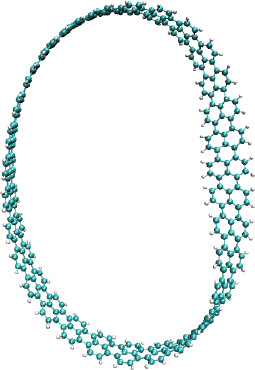}} &
\subfigure{\includegraphics[width=\sizeC]{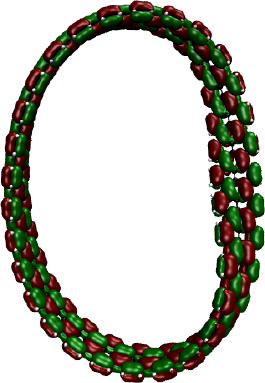}} &
\subfigure{\includegraphics[width=\sizeC]{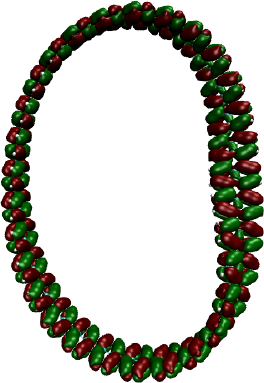}} \\

\end{tabular}
\caption{\label{FigS:OrbitalsMCNB} Frontier orbitals (HOMO and LUMO) for all
M\"obius carbon nanobelts. Top to bottom: number of repetitions from 10 to 30.}
\end{figure}
\restoregeometry

\end{document}